\documentclass[12pt,a4paper]{JHEP3}
\pdfoutput=1

\usepackage{bm}
\usepackage{mathrsfs}
\usepackage{amsmath}
\usepackage{amssymb}
\usepackage{amstext}

\usepackage{epsfig,latexsym}
\usepackage{float}
\usepackage[font=small,labelfont=bf]{caption}

\usepackage{graphicx}

\usepackage[titletoc]{appendix}

\title{Dynamical Condensation in a Holographic Superconductor Model with Anisotropy}

\author{
Xiaojian Bai $^{1,2}$, Bum-Hoon Lee $^{1,2}$, Miok Park $^{1,2}$, Khimphun Sunly$^{1,2}$\\
{}\\
{\it $^1$ Center for Quantum Spacetime, Sogang University, Seoul, Korea}\\
{\it $^2$ Department of Physics, Sogang University, Seoul, Korea}\\
\\E-mail: \email{baixj@sogang.ac.kr}, \email{bhl@sogang.ac.kr},  \email{miokpark76@gmail.com}, \email{kpslourk@sogang.ac.kr}}

\preprint{}

\newcommand{\oo}{\langle\mathcal{O}_2\rangle}
\newcommand{\boo}{\langle\overline{\mathcal{O }}_2\rangle}
\newcommand{\half}{\dfrac{1}{2}}
\newcommand{\x}{\mathbf{x}}
\newcommand{\nn}{\nonumber}
\newcommand{\qq}{\quad}
\newcommand{\tF}{\tilde{F}}
\newcommand{\tpsi}{\tilde{\psi}}
\newcommand{\talpha}{\tilde{\alpha}}
\newcommand{\tB}{\tilde{B}}
\newcommand{\tPhi}{\tilde{\Phi}}
\newcommand{\bT}{\overline{T}}
\newcommand{\dP}{\delta \mathcal{P}}

\date{\today}

\abstract{We study dynamical condensation process in a holographic superconductor model with anisotropy. The time-dependent numerical solution is constructed for the Einstein-Maxwell-dilaton theory with complex scalar in asymptotic AdS spacetime. The introduction of dilaton field generates the anisotropy in boundary spatial directions. In analogy of isotropic case, we have two black hole solutions below certain critical temperature $T_c$, the anisotropic charged black hole with and without scalar hair, corresponding respectively to the supercooled normal phase and superconducting phase in the boundary theory. 
We observe a nonlinear evolution from a supercooled anisotropic black hole without scalar hair to a anisotropic hairy black hole. Via AdS/CFT correspondence, we extract time evolution of the condensate operator, which shows an exponential growth and subsequent saturation, similar to the isotropic case. Furthermore, we obtain a nontrivial time evolution of the boundary pressure, while in isotropic case it remains a constant. We also generalize quasinormal modes calculation to anisotropic black holes and shows scalar quasinormal modes match with relaxation time scale of the condensate operator.
In addition, we present the final temperature and anisotropic pressure as functions of initial temperature and background anisotropy.
}

\begin{document}

\section{Introduction}

The AdS/CFT correspondence \cite{Maldacena:1997re, Gubser:1998bc, Witten:1998qj} has enlarged our horizon of understanding for strongly coupled quantum system over the past decade. It enables us to gain insights of an otherwise intractable quantum theory by studying its classical gravitational dual in asymptotic AdS spacetime. Many gravity models have been proposed to capture various aspects of strongly coupled many body systems. In particular, the dual model of s-wave superconductor was proposed in \cite{Hartnoll:2008vx, Hartnoll:2008kx} and dubbed as holographic superconductor. There have been many works investigating equilibrium or near-equilibrium physics related to this topic, such as theories with higher spin condensate \cite{Gubser:2008wv, Cai:2013aca, Chen:2010mk}, various correction terms \cite{Gregory:2009fj,Jing:2010zp,Gangopadhyay:2012am},  quasinormal modes \cite{quasi}, viscosity, transport properties \cite{Herzog:2008he,Herzog:2011ec,Herzog:2009md},  theories with dynamical gauge fields \cite{Domenech:2010nf} and etc. 

\begin{figure}[H]
\centering
\includegraphics[width =0.6\textwidth]{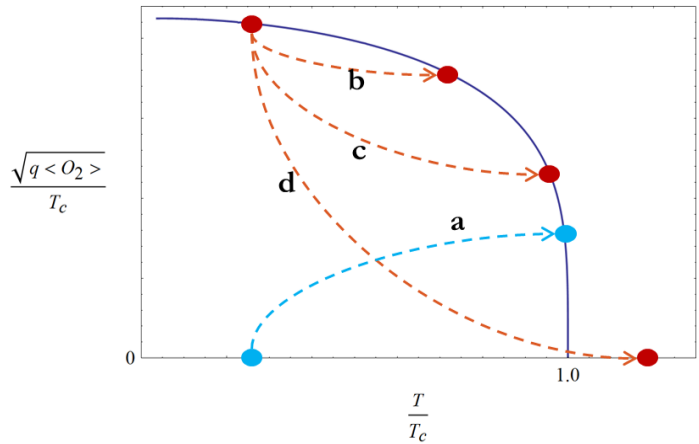} 
\caption{Cartoon picture of far-from-equilibrium dynamics of holographic superconductors. Blue route: Non-equilibium condensation from supercooled normal phase to superconducting phase. Red routes: Quantum quench of a superconducting state.}\label{phasediagram}
\end{figure}

Less work is done on studying the far-from-equilibrium physics, which on gravity side requires constructing time-dependent black hole solutions. References \cite{muruta} and \cite{wiseman} studied two basic scenarios of far-from-equilibrium  evolution of homogeneous and isotropic holographic superconductors, shown in Figure \ref{phasediagram}. The blue route (a) describes dynamical evolution from an unstable Reissner-Nordstr\"{o}m-AdS black hole to a stable hairy black hole, which is interpreted as a non-equilibrium condensation process in a holographic superconductor. The red routes (b, c and d) describe the relaxation of a superconducting state after an abrupt energy injection, a.k.a. quantum quench.  As the quench strength is increased, the condensate undergoes a dynamical phase transition from under-damped (b) to over-damped oscillation (c). At large quench strength, the system retreats to normal phase (d). Similar result is also shown in the condensed matter study \cite{barankov2006synchronization}. Other works pertaining to far-from-equilibrium dynamics of holographic superconductor include periodic driven system \cite{periodic}, thermalization with spatial inhomogeneity \cite{haiqing1} and quantum quench of a superconducting AdS soliton \cite{haiqingsoliton}. 

Recently, the s-wave holographic superconductor model is extended to incorporate spatially anisotropic effect from bulk geometry \cite{aniso}. The anisotropy is achieved by introducing a dilaton field on top of the original Einstein-Maxwell-charged scalar theory. This dilaton field does not couple with other matter fields and has a simple profile
\begin{align}
\varphi = \lambda x~,
\end{align}
where $\lambda$ is a dimensionless constant and characterizes the bulk anisotropy. 
From a phenomenological point of view, dilaton field mimic the effect of anisotropy induced by crystal structure or doping. This is because once the dilaton field is turn on, the anisotropy will sustain in both normal and superconducting phase, unlike the p-wave superconductor where the anisotropy only appears in the superconducting phase \cite{Gubser:2008wv}.
The asymptotic AdS geometry can still be preserved. In this work, we study non-equilibrium condensation process for this anisotropic holographic superconductor model. In analogy of isotropic case, there exist two static black hole solutions below certain critical temperature $T_c$, the anisotropic charged black hole with and without scalar field, corresponding respectively to the supercooled normal phase and superconducting phase of the anisotropic superconductor. The supercooled anisotropic black hole is not a stable configuration. A small scalar field perturbation will trigger a nonlinear evolution of the spacetime toward a more stable configuration, which is the anisotropic black hole with non-trivial scalar hair. The local $U(1)$ symmetry is spontaneously broken during this process. From the nonlinear dynamics of the bulk fields, we study the time evolution of the condensate operator and anisotropic pressure of the boundary system.   
In particular, note that the pressure response remains trivially
 a constant in isotropic case of non-equilibrium condensation. The introduction of dilaton is crucial to obtain a nontrivial time evolution of boundary pressure. We will discuss this point with more detail in Section \ref{sec.RG}. 

This paper is organized as follows. In Section \ref{sec.Model}, we introduce the holographic superconductor model with anisotropy and outline the numerical scheme of solving the coupled PDEs. In Section \ref{sec.RG}, we perform the holographic renormalization in time-dependent setting to obtain the renormalized stress-energy tensor and condensate operator. In Section \ref{sec.condensation}, we first recapitulate basic equilibrium results including the condensate, anisotropic pressure. Next, we present the numerical results of bulk field dynamics and the time evolution of boundary operators. The exponential growing behaviour of condensate operator is shown to match the time scale extracted from dominant quasinormal modes. In Section \ref{sec.QNMs}, we give the detail of computing quasinormal modes on anisotropic black branes with vanishing scalar hair. The conclusion and comment on future direction constitute the final section.

\section{The Model with Anisotropy}\label{sec.Model}

We consider the anisotropic holographic superconductor model in 4-dimension, described by action
\begin{align}
S=
\dfrac{1}{2\kappa^2}
\int d^4 x \sqrt{-g} \left( R +\dfrac{6}{L^2}   -\frac{1}{4} F^2 -(\nabla\varphi)^2 - |\nabla\psi - i q A \psi|^2 - m^2 |\psi|^2 \right),\label{action}
\end{align}
where 
$L$ is the AdS radius and set to $1$ hereafter. The diffeomorphism and local $U(1)$ symmetry are manifest in this action. 
$F_{\mu\nu} = \partial_{\mu}A_{\nu}-\partial_{\nu}A_{\mu}$ is the field strength of gauge field. $q$ and $m$ are respectively the charge and mass of the complex scalar field.  This action generalizes the original s-wave holographic superconductor to anisotropic background by introducing dilaton field $\varphi$. It provides a source to anisotropy in the bulk geometry and affects other matter field and condensation only through gravity. Nevertheless, the asymptotic AdS geometry of the solutions can still be preserved.

Assuming planar symmetry, we can use following ansatz in ingoing Eddington-Finkelstein coordinates,
\begin{align}
&ds^2 = - \frac{1}{z^2} \left(F(v,z)dv^2 + 2 dv dz \right) + \Phi(v,z)^2\left(e^{B(v,z)} dx^2 + e^{-B(v,z)} dy^2\right), \\
&A = \alpha(v,z)dv, \\
&\psi = \psi(v,z),\\
&\varphi = \lambda x\,.
\end{align}
The AdS boundary is at $z=0$, and $B(v,z)$ is introduced in response to the anisotropy. The dilaton field $\varphi$ is taken to be static and $\lambda$ is a dimensionless constant, which satisfies the dilaton E.O.M. automatically. This ansatz is invariant under residual symmetry 
\begin{eqnarray}
\dfrac{1}{z}\to \dfrac{1}{z}+f(v),\quad \alpha\to \alpha + \partial_v \theta(v),\quad \psi\to e^{iq\theta(v)}\psi~\,.
\end{eqnarray}

The equations of motion are 
\begin{subequations}
\label{eoms}
\begin{align}
&D\psi'+\frac{\Phi '}{\Phi }D\psi +\frac{m^2 \psi }{2 z^2}+\frac{1}{2} i q \psi  \alpha '+\frac{D\Phi}{\Phi } \psi '=0\label{eq_psi}\,,\\
&D\alpha'+ \alpha '\left(\frac{1}{2} z^2 \left(\frac{F}{z^2}\right)'+\frac{2 D\Phi }{\Phi }\right)+\frac{i  q}{z^2}\left( D\psi \psi^*-D\psi^*\psi \right)=0\label{eq_a2}\,,\\
&\alpha ''+2 \alpha '\left(\frac{1}{z}+\frac{\Phi '}{\Phi }\right)+\frac{i q}{z^2}\left( \psi  \psi^*{}'- \psi^* \psi '\right)=0\,,\label{eq_a1}\\
& (\Phi D\Phi)'-\dfrac{\Phi^2}{2z^2}\left(\frac{m^2 }{2} |\psi|^2+\frac{1}{4} z^4 (\alpha '){}^2-3\right)-\frac{e^{-B} \lambda ^2}{4 z^2}=0\,,\label{eq_phi2}\\
& \Phi '' +\frac{2 }{z}\Phi ' +\frac{1}{2} \Phi(  |\psi|'{}^2+\frac{1}{2}B'{}^2)= 0\,,\label{eq_phi1}\\
& (\Phi DB)'+(D\Phi) B'-\frac{e^{-B} \lambda ^2}{2 z^2 \Phi}=0\,,\label{eq_B}\\
& \left(z^2 \left(\frac{F}{z^2}\right)'\right)'-z^2 \left(\alpha '\right)^2+\frac{4 (D\Phi) \Phi '}{\Phi ^2}-(D\psi^* \psi '+D\psi \psi^*{}')\nn\\
&\qq-(DB) B'-\frac{e^{-B} \lambda ^2}{z^2 \Phi ^2}=0\label{eq_F}\,,\\
& D^2\Phi+\frac{1}{2} z^2 \left(\frac{F}{z^2}\right)'D\Phi+\frac{\Phi}{2} \left(|D\psi|^2 +\dfrac{1}{2}DB^2\right)=0\,,\label{eq_phi3}
\end{align}
\end{subequations}
where $'=\partial_z$ and we defined
\begin{align}
&D\Phi = \partial_{v} \Phi - F \partial_z \Phi/2,  &&D^2\Phi = \partial_v(D\Phi) - F \partial_z(D\Phi)/2,\nn\\
&DB= \partial_{v} B - F \partial_z B/2, &&D\alpha = \partial_v \alpha - F \partial_z \alpha/2,\label{Ddef}\\
&D\psi = \partial_v \psi - F \partial_z \psi/2 - iq\alpha\psi,  &&D\psi^* = \partial_v \psi^* - F \partial_z \psi^*/2 + iq\alpha\psi^*\,. \nonumber
\end{align}
The operator $\partial_v - F \partial_z /2$ represents the 
derivative along the outgoing null vector. Note that only $\lambda^2$ appears in the E.O.M., so we choose it to be positive without loss of generality. The equations in isotropic case of \cite{muruta} can be recovered by setting $\lambda=0$ and $B=0$. 

All the equations are now ordinary differential equations for either ``D" variables or original ones. The nested structure of the equations makes it possible to integrate them one at a time. Provided the initial configurations of $\psi$ and $B$ are given, one can solve \eqref{eq_a1} and \eqref{eq_phi1} to get initial condition for $\alpha$ and $\Phi$, then subsequently solve \eqref{eq_phi2}, \eqref{eq_B}, \eqref{eq_psi} and \eqref{eq_F} to obtain ``$D$" variables and further extract time derivative of original fields from \eqref{Ddef}, eventually, approximate the field value on next time slice by finite difference. An alternative approach was suggested in \cite{bible}. One may simply solve \eqref{eq_phi1} and \eqref{eq_a1} on every time slice to obtain $\Phi$ and $\alpha$, rather than evolving them dynamically through $\partial_v\Phi$ and $\partial_v\alpha$. Numerical stability can be gained from this approach. Boundary condtions are only imposed at AdS boundary $z=0$. We employ the Chebyshev Pseoduspectral method to solve ODEs at every time slice. In addtion, it is not difficult to show that if \eqref{eq_phi3} is satisfied on AdS boundary $z=0$, it will hold for all $z$ and thus will only contribute a constraint equation at the boundary.


\section{Holographic Renormalization}\label{sec.RG}
In order to interpret gravity computation in boundary field theory, we need to express gauge invariant operators and their correlation functions in terms of asymptotic coefficients of corresponding bulk fields. This procedure is achieved by holographic renormalization. Examples of performing holographic renormalization in time-dependent setting are given in \cite{Buchel:2012gw, Buchel:2013lla} which study quantum quench of holographic plasmas.

The first step is to obtain an asymptotic expansion solution of \eqref{eoms}, which depends on the mass of complex scalar field. In this work, we focus on the case of $m^2=-2$ and the expansion is 
\begin{subequations}
\label{asymp}
\begin{align}
F(v,z)& = 1+2 z \Phi _0+z^2 \left(-\frac{\lambda ^2}{2}-\frac{|\psi _1|^2}{2}+\Phi _0^2-2 \dot{\Phi} _0\right)+z^3 F_3+\mathcal{O}(z^4)\,,\\
\Phi(v,z)& =\dfrac{1}{z}+\Phi _0-\frac{1}{4} z |\psi _1|^2-\frac{1}{12} z^2 \left(2 \left(\psi _2 \psi^*_1+\psi _1 \psi^*_2\right)+\Phi _0 |\psi _1|^2\right)+\mathcal{O}(z^3)\,,\\
B(v,z)& = \frac{z^2 \lambda ^2}{2}+z^3 B_3+\frac{1}{4} z^4 \left(\lambda ^4+\lambda ^2|\psi _1|^2+4 \dot{B}_3\right)\notag\\
&+\frac{1}{4} z^4 \left(-12 B_3 \Phi _0-6 \lambda ^2 \Phi _0^2+4 \lambda ^2 \dot{\Phi} _0\right)+\mathcal{O}(z^5)\,,\\
\psi(v,z)& = z \psi _1+z^2 \psi _2+z^3 \left(-\frac{1}{2} i q \alpha _1 \psi _1-i q \alpha _0 \psi _2+\frac{1}{2} \psi _1^2 \psi^*_1+\dot{\psi}_2\right)\notag\\
&+z^3 \left(-\Phi _0^2 \psi _1+\psi _1 \dot{\Phi} _0+\Phi _0 \left(-i q \alpha _0 \psi _1-2 \psi _2+\dot{\psi} _1\right)\right)+\mathcal{O}(z^4)\,,\\
\alpha(v,z)& = \alpha _0+z \alpha _1+\frac{1}{2} z^2 \left(i q \left(\psi _2 \psi^*_1-\psi _1 \psi^*_2\right)-2 \alpha _1 \Phi _0\right)+\mathcal{O}(z^3)\,,
\end{align}
\end{subequations}
where the asymptotic AdS condition has been imposed, $F\rightarrow 1$, $\Phi \rightarrow 1/z$ and $B\rightarrow0$ as $z\rightarrow 0$.
In addition, we have two constraints at the boundary 
\begin{subequations}
\label{constraints}
\begin{align}
\dot{F}_3(v)&=\frac{1}{2} \psi_1^*D_v\left(-\psi_2-\Phi _0 \psi _1 +D_v\psi_1\right) +c.c.\,,\label{bdF3}\\
\dot{\alpha} _1(v)&=i q \psi _1^* \left(\psi_2-D_v\psi_1\right)+c.c. \,,\label{bda1}
\end{align}
\end{subequations}
where $D_v \equiv\partial_v - iq\alpha_0$ and $c.c.$ is complex conjugate. We identify  $\alpha_0$ and $\alpha_1$ as chemical potential and charge density of the boundary theory. $B_3$ and $F_3$ are related to anisotropic pressure and energy density. $\psi_1$ is the sourse and $\psi_2$ the vacuum expectation value.

The residual diffeomorphism 
\begin{align}
\dfrac{1}{z}\to \dfrac{1}{z}+f(v)
\end{align}
transform boundary coefficients 
\begin{equation}
\left( \begin{array}{c}
\psi_2 \\
\Phi_0\\
B_3
\end{array} \right)
\to \left( \begin{array}{c}
\psi _2-\psi _1 f\\
\Phi_0+f\\
B_3 + \lambda^2 f
\end{array} \right) \,,
\end{equation}
and leave $\psi_1,\alpha_0,\alpha_1$ and $F_4$ unchanged. We use this symmetry to set $\Phi_0=0$ for later numerical calculation. $\alpha_0$ can also be set to zero by the residual $U(1)$ symmetry.

We transform the solution to Fefferman-Graham(FG) coordinate (denoted by $g_{\mu\nu}$) where the holographic renormalization can be most conveniently implemented
\begin{align}
ds^2_4 = \dfrac{d\rho^2}{\rho^2}+\dfrac{1}{\rho^2}G_{ij}(\bold{x},\rho)dx^{i}dx^{j},\quad \psi=\psi(\bold{x},\rho),\quad A=\alpha(\bold{x},\rho)dt\,.
\end{align}
The asymptotic form of coordinate transformation is
\begin{align}
v &= t-\rho -\frac{1}{24} \rho ^3 \left(\lambda ^2+|\psi _1|^2 \right)+\frac{1}{96} \rho ^4 \left(8 F_3+\dot{\psi}^*_1 \psi _1+\dot{\psi} _1 \psi^*_1\right) + \mathcal{O}(\rho^5)\,,\\
z &= \rho +\rho^2 \Phi_0 -\frac{1}{8} \rho ^3 \left(\lambda ^2+|\psi _1|^2- 8\Phi _0^2+8\dot{\Phi} _0 \right)+\frac{1}{12} \rho ^4 \left(2 F_3+\dot{\psi}^*_1 \psi _1+\dot{\psi} _1 \psi^*_1\right)\notag\\
&\quad +\frac{1}{4} \rho ^4 \left(4 \Phi _0^3-8\Phi _0 \dot{\Phi} _0+2 \ddot{\Phi} _0-\Phi _0 \left(\lambda ^2+|\psi _1|^2\right)\right)+\mathcal{O}(\rho^5)\,,
\end{align}
and it is easy to see that $v$ and $t$ coincide as one approaching the AdS boundary $\rho=0$. Note this coordinate transformation necessarily introduces a $A_{\rho}$ component which should be removed through $U(1)$ transformation, in accordance with our radial gauge. Scalar field $\psi$ therefore will pick up a non-trivial phase factor.

The second step is to identify the divergent pieces and construct counterterms.
The Einstein equation implies
\begin{align}
E_{xx}= \half g_{xx}\left( \mathcal{L} - R \right)+\lambda^2\,,\quad
E_{yy}= \half g_{yy}\left( \mathcal{L} - R \right)\,,
\end{align}
where $E_{ab}$ is the Einstein tensor, therefore
\begin{align}
\mathcal{L} = -E^t{}_t-E^z{}_z-\lambda^2 g^{xx}\,.
\end{align}
Using this, the on-shell action can be written as
\begin{align}
S_{\text{on-shell}} &=\dfrac{1}{2\kappa^2} \int d^2x\,dt\,\int^{\epsilon}_{\rho_+}d\rho\left(- \partial_\rho\left( \rho\sqrt{-g_{tt}}\partial_\rho\left( \sqrt{g_{xx}g_{yy}} \right) \right)-\dfrac{\lambda^2}{\rho}\sqrt{\dfrac{g_{tt}g_{yy}}{g_{xx}}}\right)\notag\\
&+\dfrac{1}{2\kappa^2}\int d^2x\,d\rho\int^{+\infty}_{-\infty}\,dt\,\partial_t\left(\dfrac{1}{\rho\sqrt{-g_{tt}}} \partial_t\left(\sqrt{g_{xx}g_{yy}} \right) \right)\,.
\end{align}
The surface term in time direction is zero, since we assume dynamics only happens in finite period of time and spacetime is static at infinity past and future.
One can show that the following counter terms will render finite results
\begin{align}
S_{c.t.}=\dfrac{1}{2\kappa^2}\int_{\rho=\epsilon} d^3x\sqrt{-\gamma}\left( 2K+4 +\psi^2 - \partial_i\varphi\partial^i\varphi\right)\,,
\end{align}
where $\gamma_{ij}= G_{ij}/\rho^2$ and $K$ is the trace of extrinsic curvature. The divergence stemmed from $-\frac{\lambda^2}{\rho}\sqrt{\frac{g_{tt}g_{yy}}{g_{xx}}}$ is cancelled by $\partial_i\varphi\partial^i\varphi$.

The last step is to invoke AdS/CFT correspondence 
\begin{align}
&\langle \mathcal{T}_{ij}\rangle 
= \dfrac{-2}{\sqrt{G_{(0)}}}\dfrac{\delta S_{ren}}{\delta G^{ij}_{(0)}}\nn\\
&=\dfrac{1}{2\kappa^2}\lim_{\rho\rightarrow0}\dfrac{-2}{\rho}\left(K_{ij}- K\gamma_{ij} -2\gamma_{ij}-\half \psi^2\gamma_{ij}+\half \partial_k\varphi\partial^k\varphi \gamma_{ij}  -\partial_i\varphi\partial_j\varphi \right)\,,\\
&\oo 
=\dfrac{1}{\sqrt{G_{(0)}}}\dfrac{\delta S_{ren}}{\delta \psi_1^*}
= \dfrac{1}{2\kappa^2}\lim_{\rho\rightarrow0}\dfrac{1}{\rho^2}\left(-n_{\rho}\partial^{\rho}\psi + \psi \right)\,,\\
&\langle{\mathcal{J}}^t\rangle 
 =\dfrac{1}{\sqrt{G_{(0)}}}\dfrac{\delta S_{ren}}{\delta \alpha_0}
=\dfrac{1}{2\kappa^2}\lim_{\rho\rightarrow0}\dfrac{1}{\rho^3} \left(- n_{\rho}F^{\rho t} \right)\,,
\end{align}
and $S_{ren}=S+S_{c.t.}$ and $G_{(0)}^{ij}=G^{ij}(\x,\epsilon)\vert_{\epsilon\rightarrow0}$. The renormalized one-point functions 
\begin{align}
\langle \mathcal{T}_{ij}\rangle\equiv\left( \mathcal{E},\mathcal{P}_x,\mathcal{P}_y \right),
\end{align}
$\langle \mathcal{O}_2\rangle$ and $\langle\mathcal{J}^t\rangle$
are
\begin{align} 
&2\kappa^2\mathcal{E} = 2 F_3 +\left( \psi^*_1\left( \psi _2 +\Phi _0 \psi _1 -D_t\psi _1 \right) +c.c. \right)\,,\\
&2\kappa^2\mathcal{P}_x =F_3-3 B_3-3 \lambda ^2 \Phi _0\,,\label{Px}\\
&2\kappa^2\mathcal{P}_y =F_3+3 B_3+3 \lambda ^2 \Phi _0\,,\label{Py}\\
&2\kappa^2\oo = -\psi _2-\Phi _0 \psi _1  +D_t\psi _1\,,\\
&2\kappa^2\langle{\mathcal{J}}^t\rangle = \alpha_1 - \dot{\alpha}_0\,,
\end{align}
where $D_t\equiv\partial_t-i q \alpha_0$ and $c.c.$ denotes complex conjugate. 
The nontrivial component of diffeomorphism Ward identity is
\begin{align}
\partial^t\langle \mathcal{T}_{tt}\rangle=-\partial_t\mathcal{E} =\oo\left( D_t\psi_1\right)^* +c.c.\,,
\end{align}
which is the boundary energy conservation. A time-dependent source term $\psi_1(t)$ will bring energy variation in the system. The conformal Ward identity is given by
\begin{align}
\langle \mathcal{T}^i{}_{i}\rangle = \langle\mathcal{O} _2\rangle\psi^*_1+ c.c.\,,
\end{align}
which carries the overall factor $(d-\Delta)=1$. $d$ is boundary spacetime dimension and $\Delta$ is the conformal dimension. The anisotropic pressure is defined as
\begin{eqnarray}
\delta \mathcal{P} = 2\kappa^2(\mathcal{P}_x - \mathcal{P}_y)~.
\end{eqnarray}

From \eqref{Px} and \eqref{Py}, we can see that in the isotropic limit where $\lambda$ and $B_3$ are zero, the boundary pressure is just $F_3$. And from \eqref{bdF3}, fixing boundary condition $\psi_1=0$ implies $F_3$ is a constant, therefore the pressure is also a constant.  By introducing dilaton field and generating bulk anisotropy, the pressure can have a nontrivial time evolution due to $B_3$.


\section{Condensation on Anisotropic Background}\label{sec.condensation}

In this section, we begin with a recap of equilibrium results which provide the initial condition for later non-equilibrium evolution of the system. Next, we demonstrate the dynamics of bulk fields after releasing two initial wave packets, and an interesting scattering is observed in the early time evolution. Finally, we present dynamics of boundary operators including condensate and anisotropic pressure. The relaxation time scale from nonlinear evolution is shown to have a good match with dominant scalar quasinormal modes.
The evolution of event and apparent horizon will be presented elsewhere along with the holographic entanglement entropy probe.

\subsection{Recap of Equilibrium Results}
The action \eqref{action} admits the static anisotropic black hole solution with complex scalar field identically vanishing, corresponding to normal phase of holographic superconductor. This solution exists for all temperature $T$ 
\footnote{Black hole temperature is defined as usual 
\begin{equation}
 T=\left.-\frac{1}{4\pi}\frac{dF}{dz}\right|_{z=z_+},\quad F(z_+)=0\,.\nn
\end{equation}}. The numerical solution was constructed in \cite{Iizuka:2012wt}, as well as a perturbative solution in small $\lambda$. If the dilaton field is absent, it just reduces to the Reissner-Nordstr\"{o}m-AdS black hole. Below critical temperature $T_c$, there is another black hole solution with non-trivial scalar field configuration, which corresponds to the superconducting phase. 

The equilibrium equations can be obtained by setting $v$-derivative to zero in \eqref{eoms}. For the convenience of numerical computation, we fix the horizon at $z=1$ and make transformation $\psi\rightarrow z\psi(z)$ and $\Phi \rightarrow \Phi(z)/z$. Then, the expansion solution near horizon is
\begin{eqnarray}
\psi(z)&=&\tpsi _0-\frac{2 \tpsi _0}{F_1}(z-1)+...\\
\alpha(z)&=&\talpha _1 (z-1)+\frac{\talpha _1 }{4\tF_1}\left(-4 \tF_1+\talpha _1^2+\frac{2 e^{-\tB_0} \lambda ^2}{\tPhi _0^2}\right.\\
&&\left.-4 \left(3-\left(-1+q^2\right) \tpsi _0^2\right)\right)(z-1)^2+...
\\
B(z)&=&\tB_0-\frac{e^{-\tB_0}\lambda ^2}{\tF_1 \tPhi _0^2} (z-1) +...\\
\Phi(z) &=& \tPhi _0+\frac{e^{-\tB_0}}{4 \tF_1 \tPhi _0}\left(-2 \lambda ^2+e^{\tB_0} \tPhi _0^2 \left(12+4 \tF_1-\talpha _1^2+4 \tpsi _0^2\right)\right) (z-1) +...\\
F(z)&=& \tF_1 (z-1)+\frac{1}{2} \left(2 \tF_1+\talpha _1^2+\frac{e^{-\tB_0} \lambda ^2}{\tPhi _0^2}\right) (z-1)^2+...
\end{eqnarray}
where $F(1)=0$ and $\alpha(1)=0$ have been imposed. The former one is a condition for event horizon, the latter one is to render $A^{\mu}A_{\mu}$ finite. There are five free parameters at horizon, $\tpsi_0,~\talpha_1,~\tB_0,~\tPhi_0$ and $\tF_1$. It's straightforward to vary one of them and shoot the rest four to match conditions on AdS boundary,
\begin{eqnarray}
B(0)=0,\quad \Phi(0)=1,\quad F'(0)=0,\quad \psi(0)=0~.\label{adscon}
\end{eqnarray}

Working in the canonical ensemble, we introduce following normalizaiton for condensate operator and temperature
\begin{eqnarray}
\langle\overline{\mathcal{O }}_2\rangle= 2\kappa^2 \dfrac{\sqrt{2}}{\alpha_1}\oo, \quad \overline{T} = \dfrac{T}{\sqrt{\alpha_1}}~\,,
\end{eqnarray}
where $\alpha_1$ is boundary charge density.

\begin{figure}[H]
\centering
\includegraphics[width =0.46\textwidth]{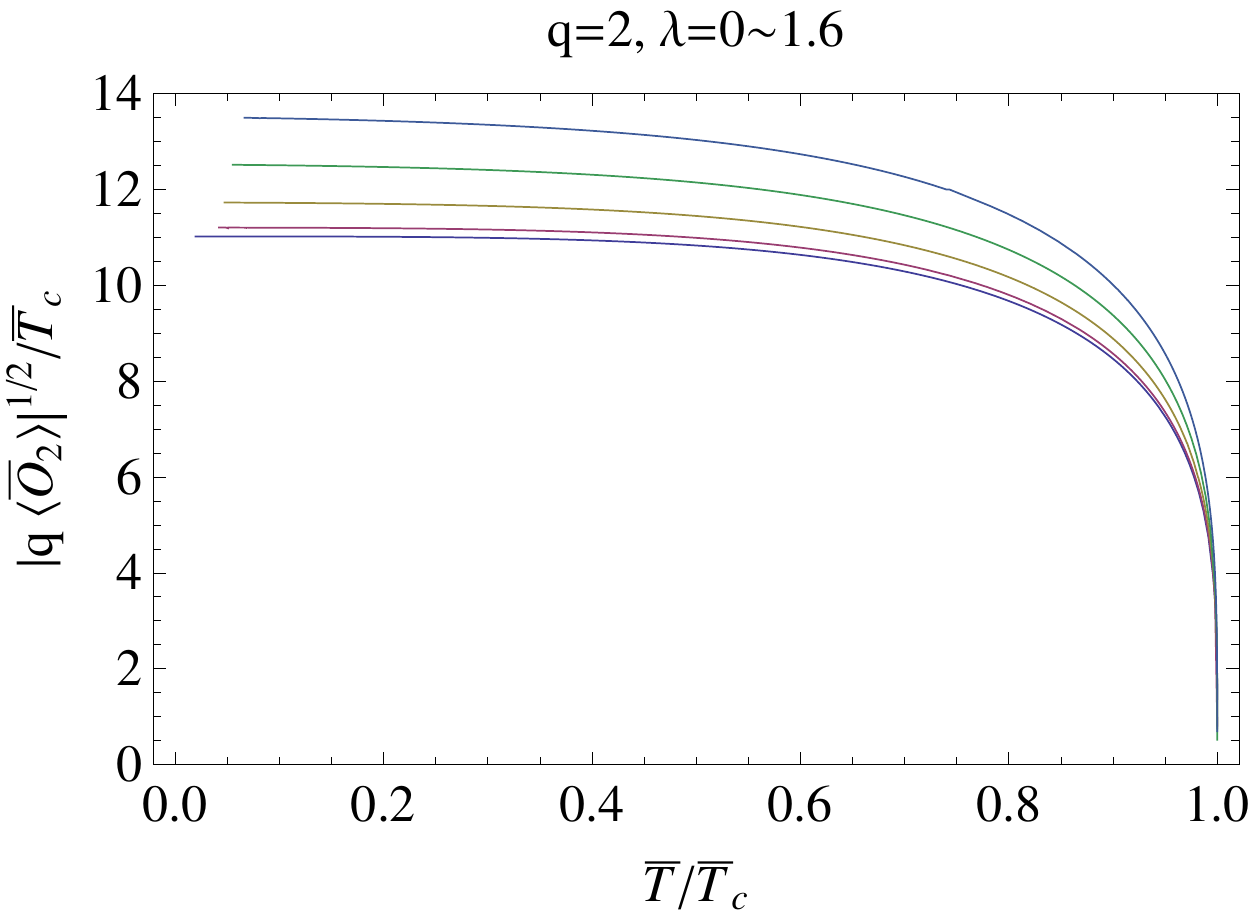} 
\hspace{0\textwidth}
\includegraphics[width =0.48\textwidth]{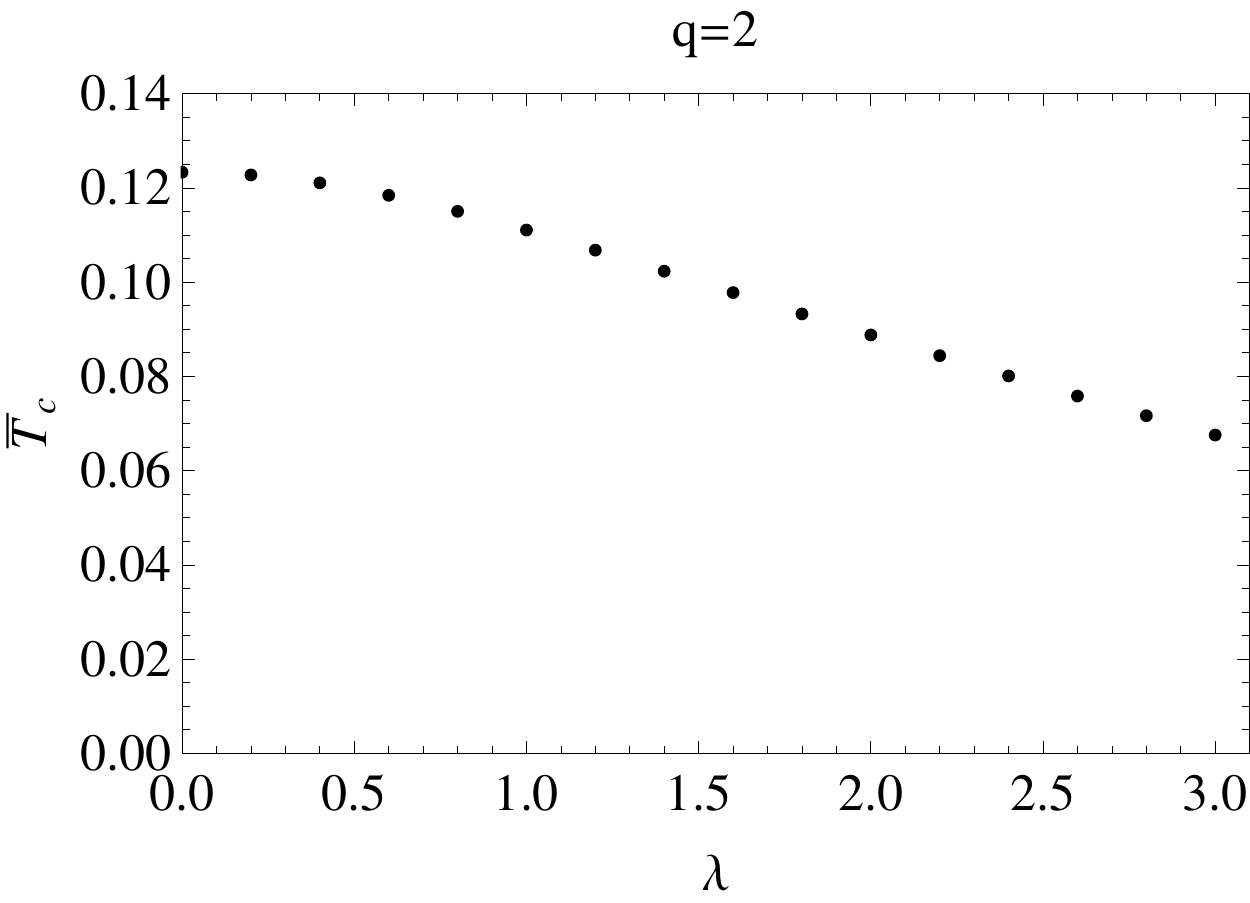} 
\caption{Left: The value of the condensate as a function of temperature with $q = 2$ for different anisotropic parameter $\lambda$. From bottom to top, various curves correspond to $\lambda = 0, 0.4, 0.8, 1.2$ and $ 1.6 $. Right: The critical temperature as a function of anisotropic parameter $\lambda$. }\label{equilcondensate}
\end{figure}

\begin{figure}[H]
\centering
\includegraphics[width =0.46\textwidth]{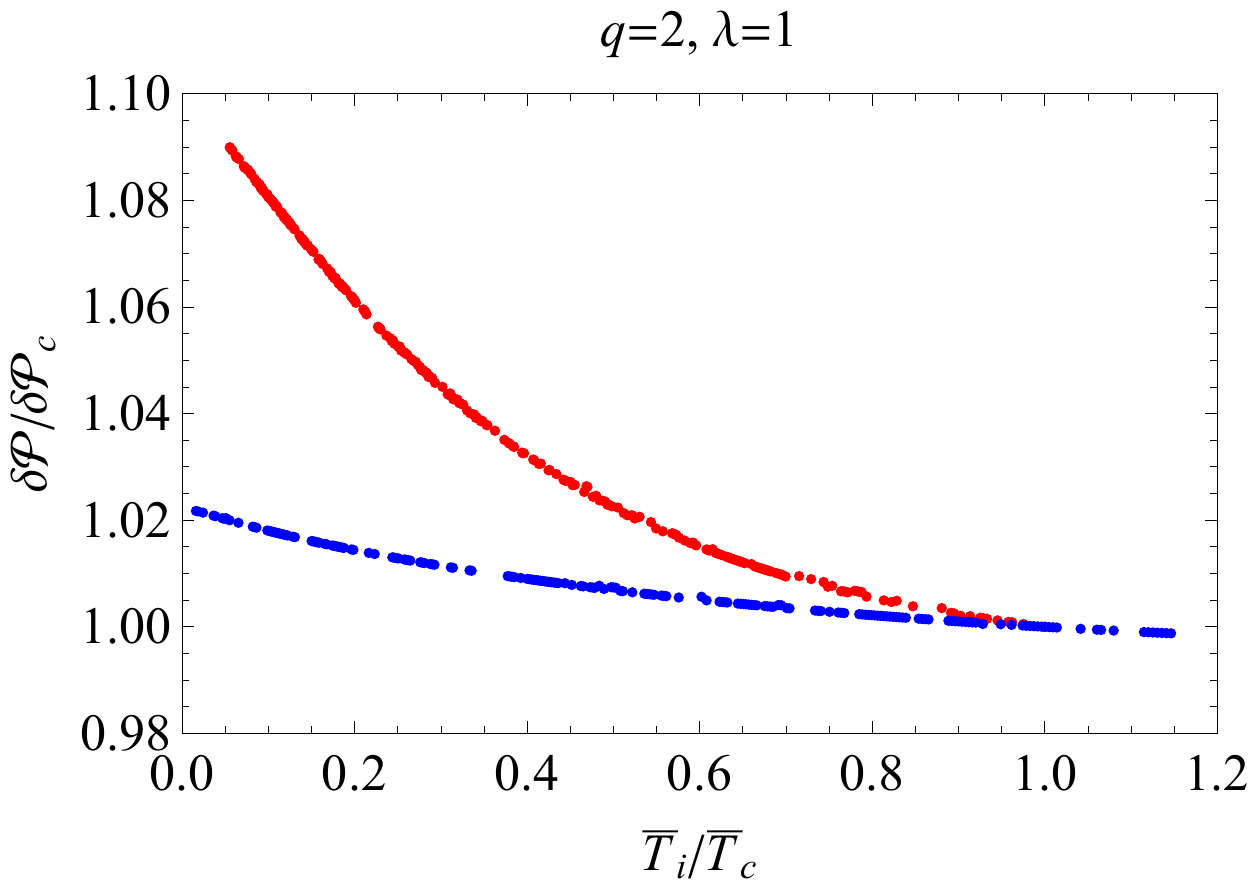} 
\hspace{0\textwidth}
\includegraphics[width =0.46\textwidth]{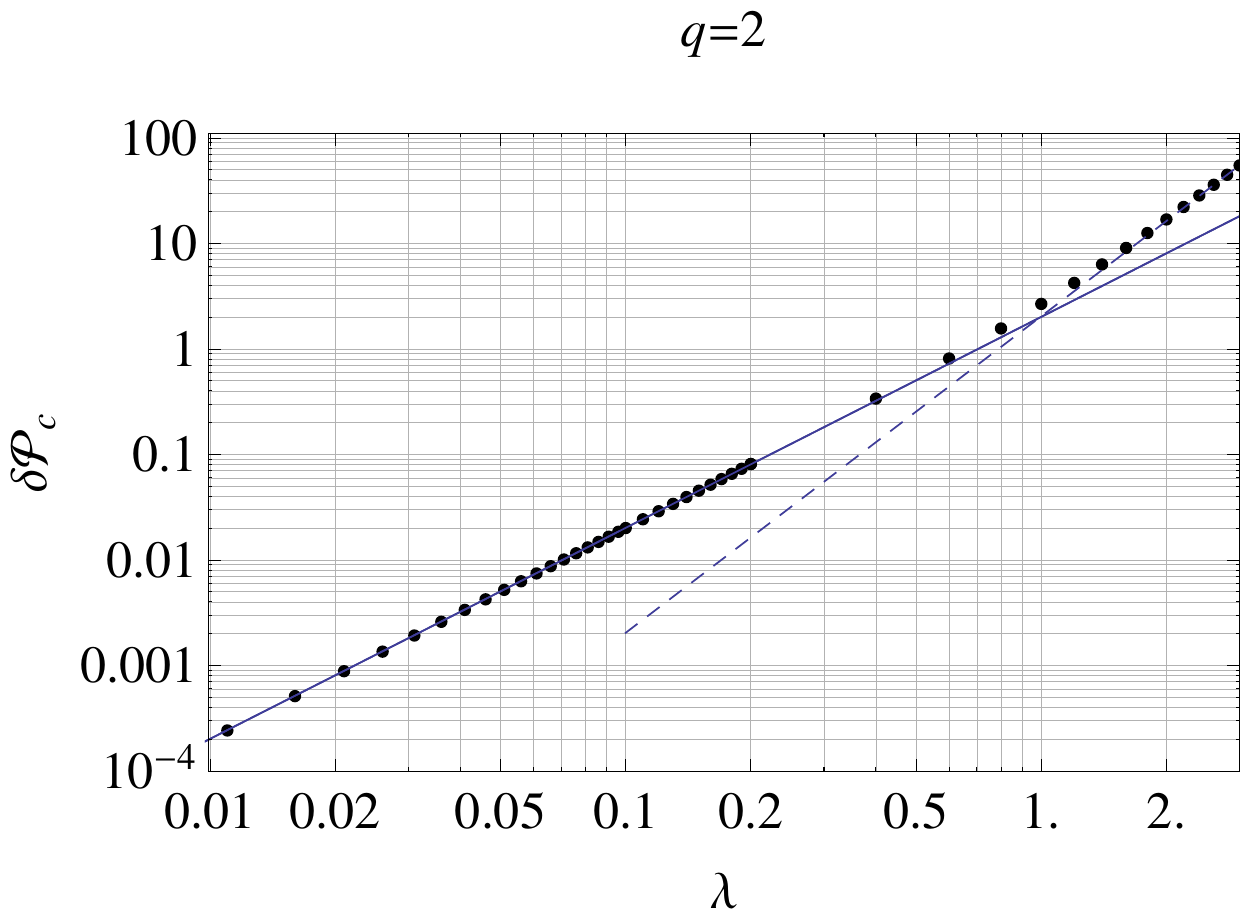} 
\caption{Left: The value of the anisotropic pressure as a function of temperature with $q=2$ and $\lambda=1$. The blue data is from anisotropic black branes with vanishing scalar hair. The red data is from anisotropic hairy solutions. Right: The anisotropic pressure of the critical point as a function of $\lambda$. The solid line is $\lambda^2$ fitting and the dashed $\lambda^3$ fitting. }\label{equilpressure}
\end{figure}

Figure \ref{equilcondensate} shows effect of bulk anisotropy on the condensate, which is also obtained in \cite{aniso}. The numerical result suggests the increasing anisotropy $\lambda$ in the bulk lifts up the condensate in the boundary for a fixed temperature, whereas brings down critical temperature $\overline{T}_c$. Whether this trend persists for large $\lambda$ is yet to be clarified. In the later non-equilibrium study, we only focus on the range $0<\lambda<3$. Figure \ref{equilpressure} shows the anisotropic pressure as a function of temperature and $\lambda$. The left plot implies the superconducting state bears bigger anisotropic pressure than the normal state and this difference vanishes continuously as the critical temperature is reached. On the right, we plot the anisotropic pressure of the critical point $\dP_c$ as a function of $\lambda$. We found $\dP_c \propto \lambda^2$ for small $\lambda$, and the dependence is enhanced to $\lambda^3$ as bulk anisotropy gets larger.

\subsection{Evolution of Bulk Fields}

First, let us summarize the initial and boundary conditions needed for numerical calculation.
The general procedure of solving \eqref{eoms} is outlined at the end of Section \ref{sec.Model}. We  use residual symmetry to fix $\Phi_1=0$.\footnote{$\alpha_0$ is kept constant throughout computation.} Since we are only concerned with condensation process, $\psi_1=0$ is chosen at AdS boundary.\footnote{The source $\psi_1$ can choose to be a Gaussian-type function, if one studies quantum quench process. Correspondingly, $F_3$ and $\alpha_1$ should be updated by boundary constraints at every time slice, see \cite{wiseman}.} It follows that $F_3$ and $\alpha_1$ are constants throughout computation, see \eqref{constraints}. All the other boundary conditions can be derived from asymptotic expansion \eqref{asymp}. The initial condition is the anisotropic charged black brane without scalar hair constructed numerically in previous subsection, along with a small scalar perturbation of following type
\begin{eqnarray}
\psi_{\text{perturb}} = \dfrac{a}{\sqrt{2\pi}\delta}z^2\exp\left( -\dfrac{(z-z_{\text{max}})^2}{2\delta^2} \right)
\end{eqnarray}
A field redefinition prior to numerical calculation is introduced to isolate the terms that only contain $\psi_1$ and divergent pieces in bulk fields. References \cite{muruta} and \cite{bible} explain this procedure quite thoroughly, we do not repeat the details here.

The general feature of scalar field $|\psi(v,z)|$ evolution is qualitatively the same as isotropic case presented in \cite{muruta}. The wave packet gets absorbed into black hole after bouncing back from the AdS boundary. Surviving modes keep growing exponentially until saturation is reached. We found the exact saturation time depends on specific parameters ($a, \delta$ and $z_{\text{max}}$) of initial perturbation, while the relaxation time scale during exponential growing period is only controlled by the dominant quasinormal modes. In response to condensation of scalar field, anisotropic function $B(v,z)$ also undergoes a dynamical change.

\begin{figure}[H]
\centering
\includegraphics[width =0.45\textwidth]{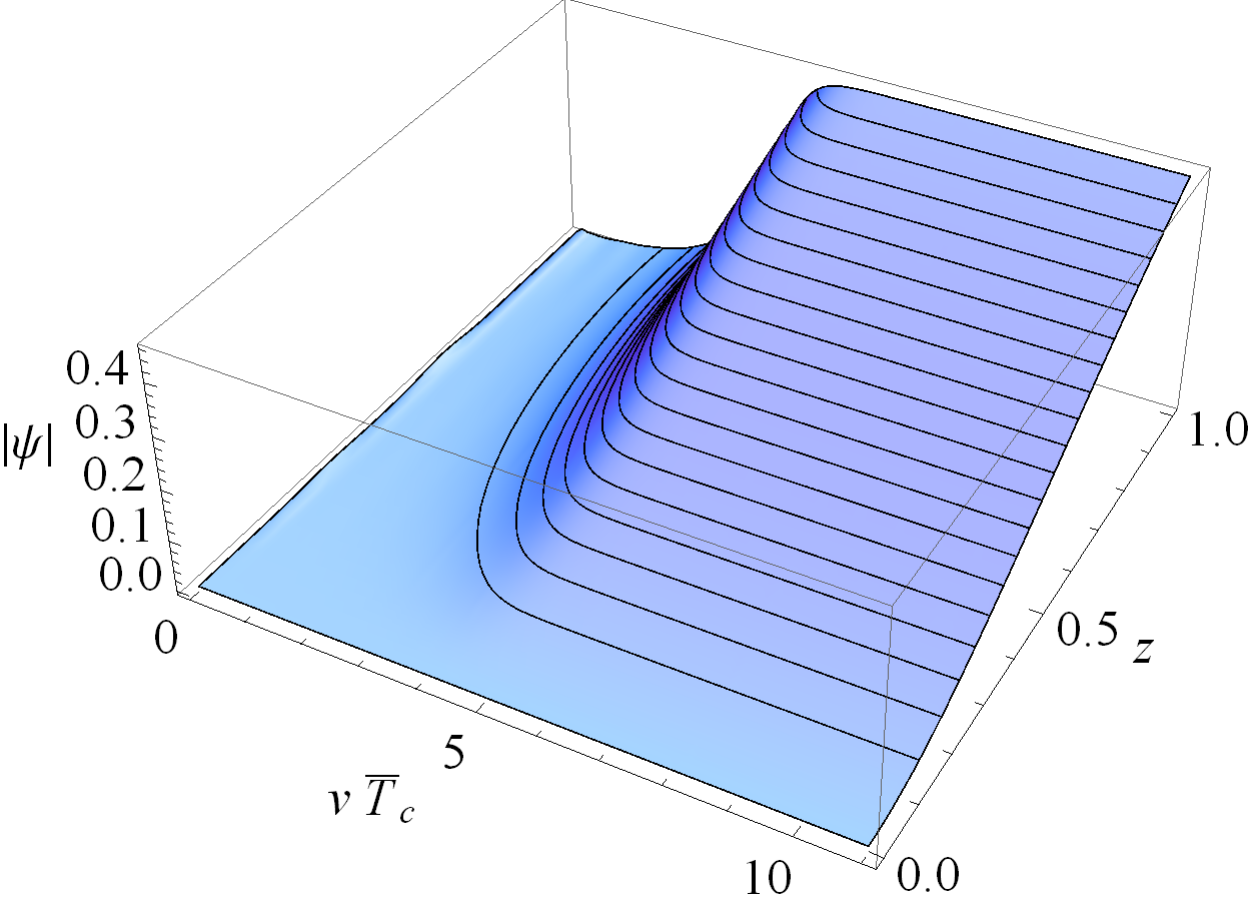} 
\hspace{0\textwidth}
\includegraphics[width = 0.45\textwidth]{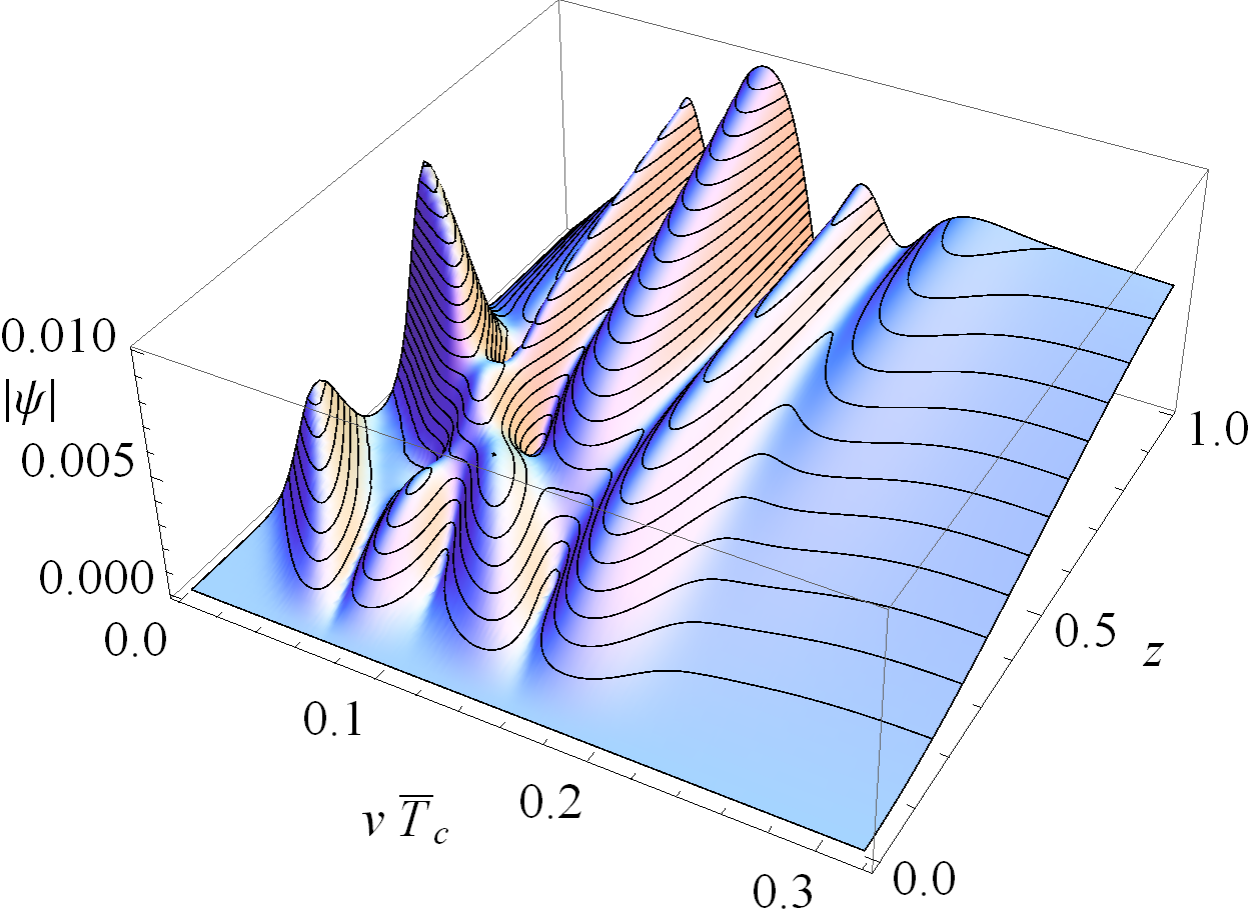} 
\caption{The evolution of scalar field $|\psi(v,z)|$ for $q=2$, $\lambda=1$ and initial temperature $\bT_i/\bT_c=0.3$. Left: The full evolution of $|\psi(v,z)|$. The scalar field undergoes an exponential growth ($0\leq v \bT_c \leq 5$) and subsequent saturation ($v \bT_c \approx 5$). Right: The early time behaviour of $|\psi(v,z)|$. The first wave packet is bounced back from AdS boundary  ($v \bT_c \approx 0.07$) and scattered by the second packet. }\label{psi3D}
\end{figure}

\begin{figure}[H]
\centering
\includegraphics[width =0.45\textwidth]{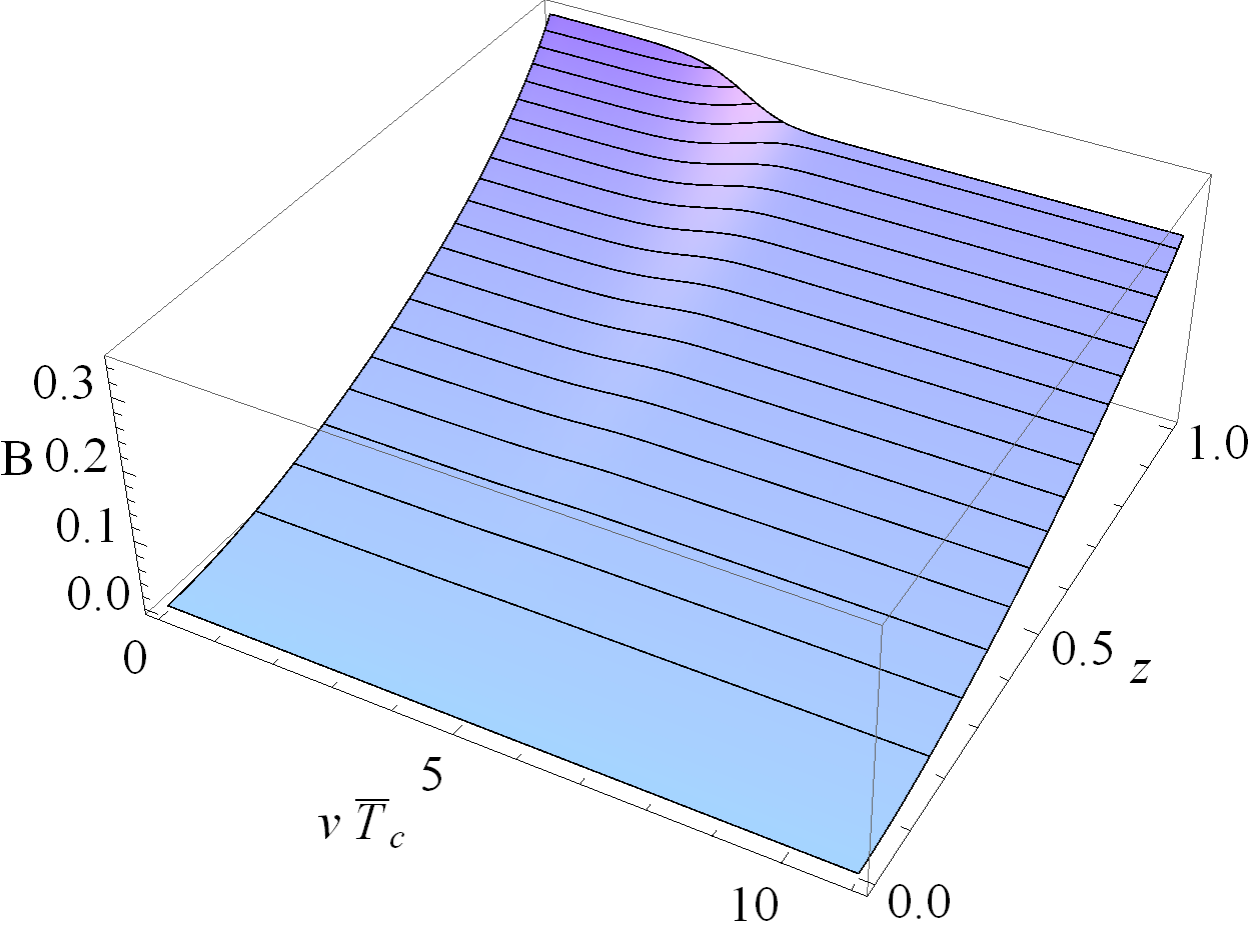} 
\hspace{0\textwidth}
\includegraphics[width = 0.45\textwidth]{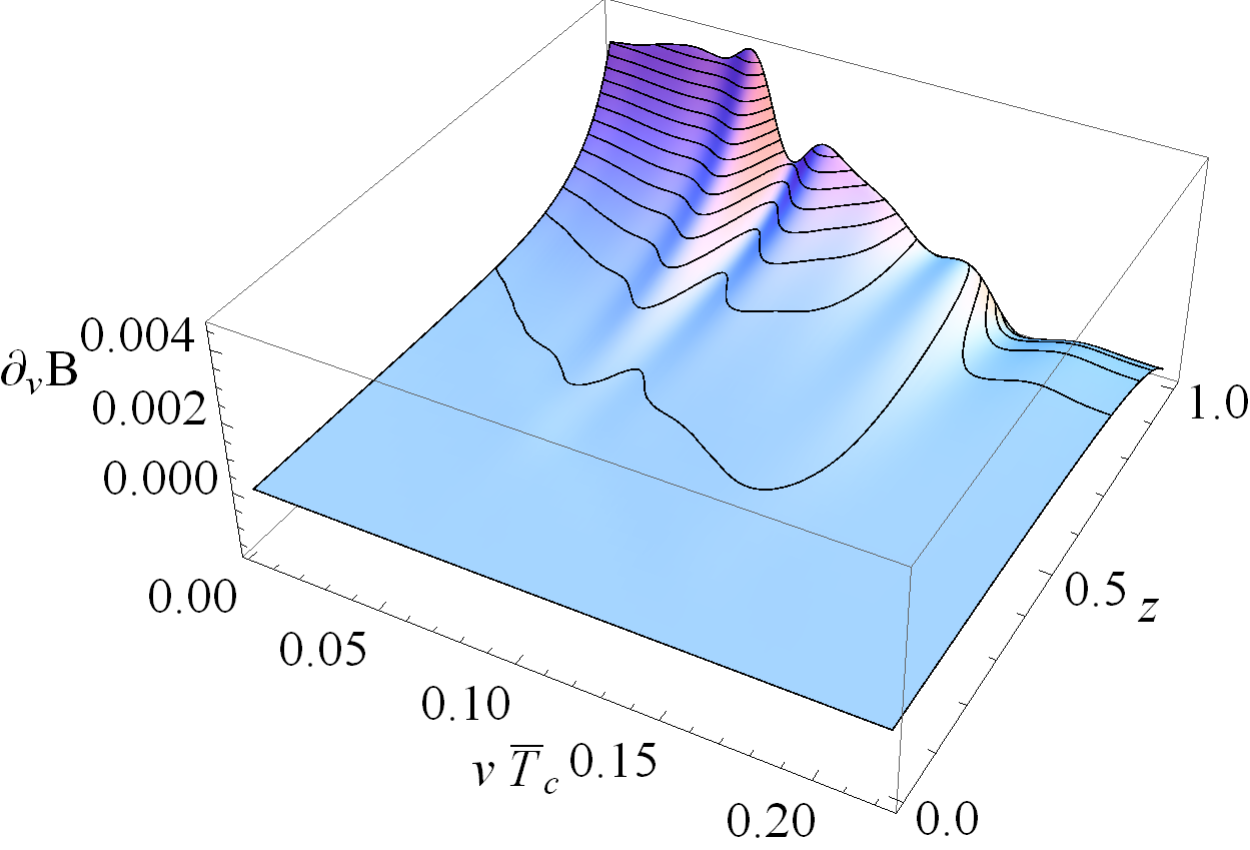} 
\caption{The evolution of anisotropic function $B(v,z)$ for $q=2$, $\lambda=1$ and initial temperature $\bT_i/\bT_c=0.3$. Left: The full evolution of $B(v,z)$. Right: The early time behaviour of $\partial_v B(v,z)$.}  \label{B3D}
\end{figure}

Figure \ref{psi3D} and \ref{B3D} show the dynamics of $|\psi(v,z)|$ and $B(v,z)$ for $q=2$ and $\lambda=1$ with initial temperature $\bT_i/\bT_c=0.3$, beginning with two initial wave packet situated at $z_{\text{max}} = 0.3$ and $z_{\text{max}} =0.6$. Two packets propagate separately toward the AdS boundary until around $v \bT_c \approx 0.07$. The reflection of first wave packet is scattered by the second one into two components which are quickly absorbed by black hole. After the scattering, the second wave packet keeps moving towards boundary and is reflected at $v \bT_c \approx 1.6$. The early time response of $B(v,z)$ to the scalar perturbation is quite small and hard to discern, so we show the plot of $\partial_v B(v,z)$ instead of $B(v,z)$.

\subsection{Evolution of Boundary Operators}
\begin{figure}[H]
\centering
\includegraphics[width =0.46\textwidth]{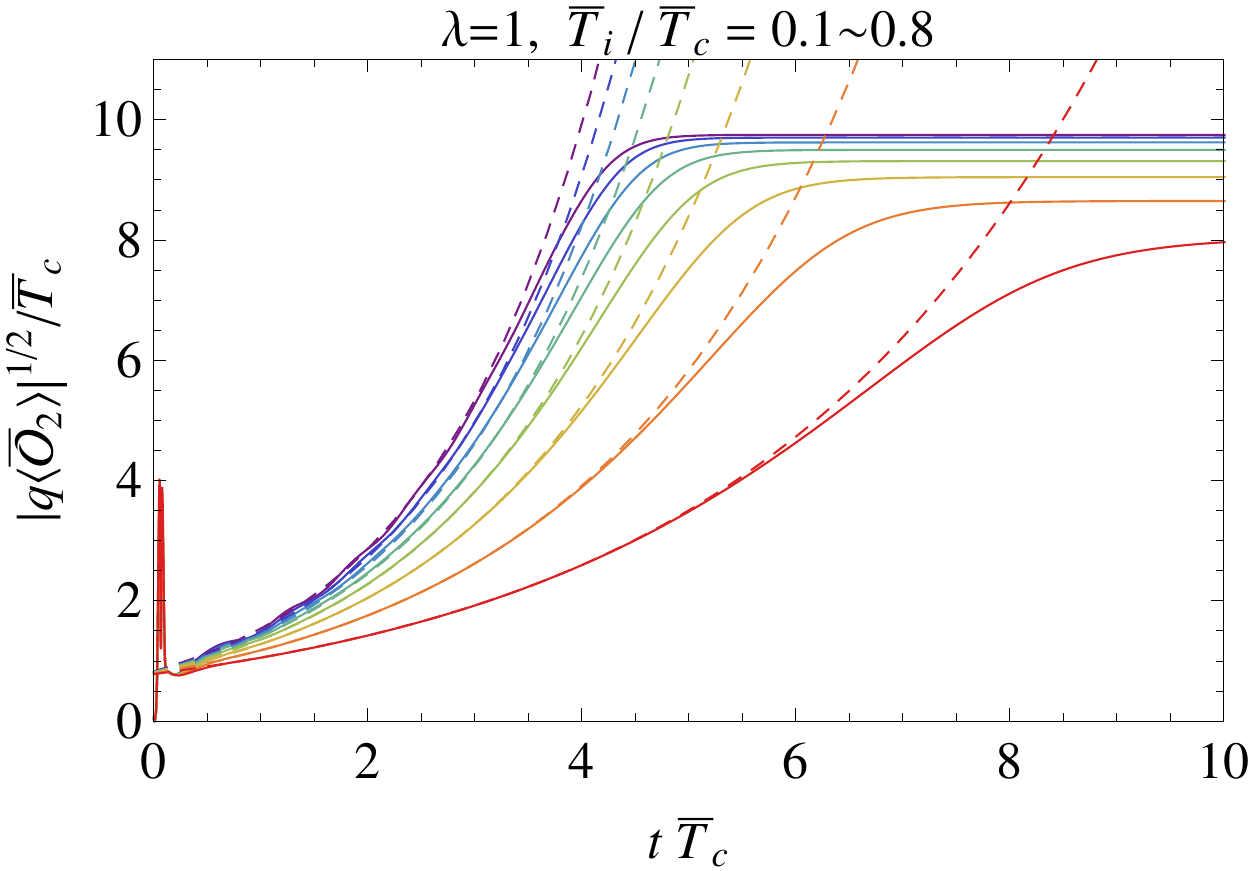} 
\hspace{0\textwidth}
\includegraphics[width = 0.46\textwidth]{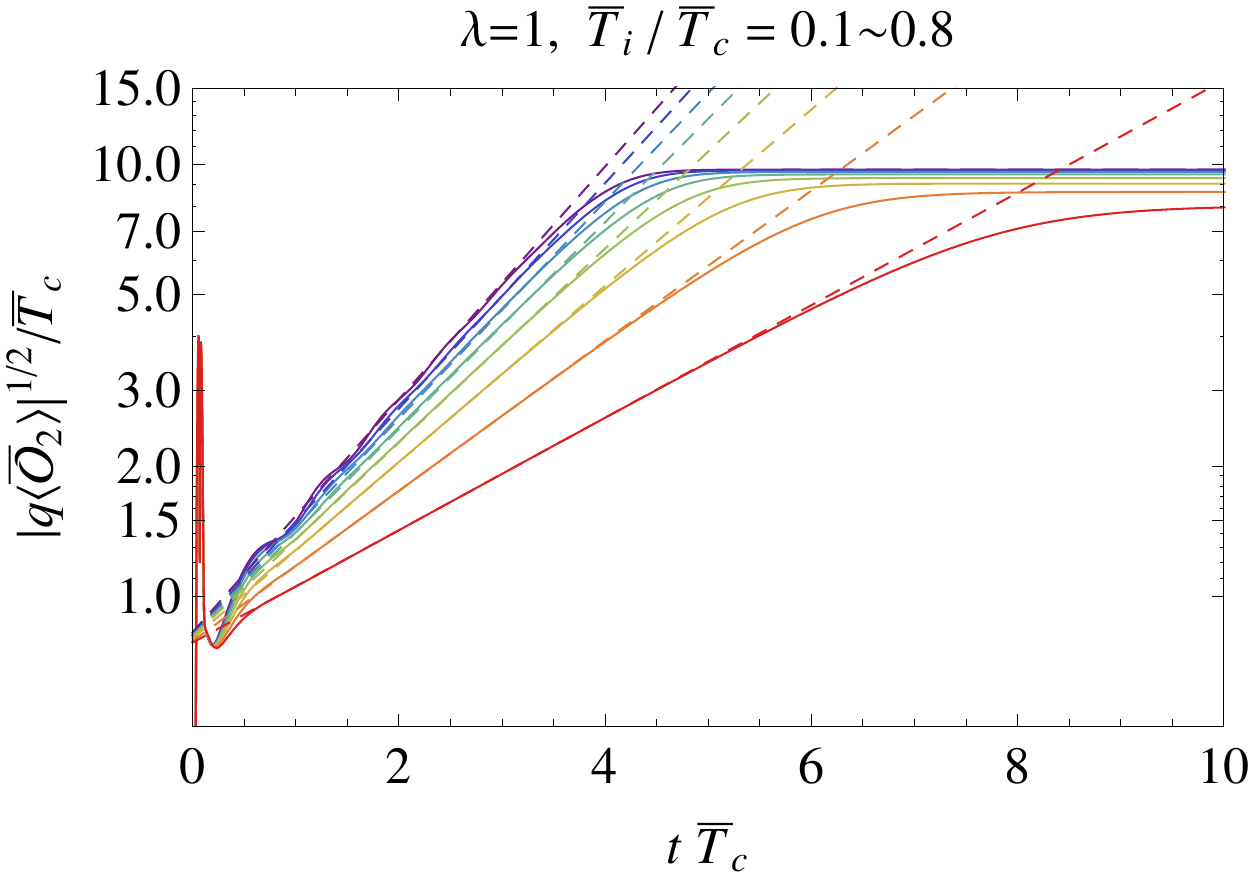} 
\caption{Left: The time evolution of condensate operator for $q=2$ and $\lambda = 1$. The solid curves from top to bottom correspond to $\bT_i/\bT_c = 0.1\sim 0.8$ with increment 0.1. The dashed curves represent the relaxation time scale obtained from leading quasinormal modes at corresponding temperature. Right: The logrithmic plot of the left.}\label{O2belowTc}
\end{figure}

\begin{figure}[H]
\centering
\includegraphics[width =0.46\textwidth]{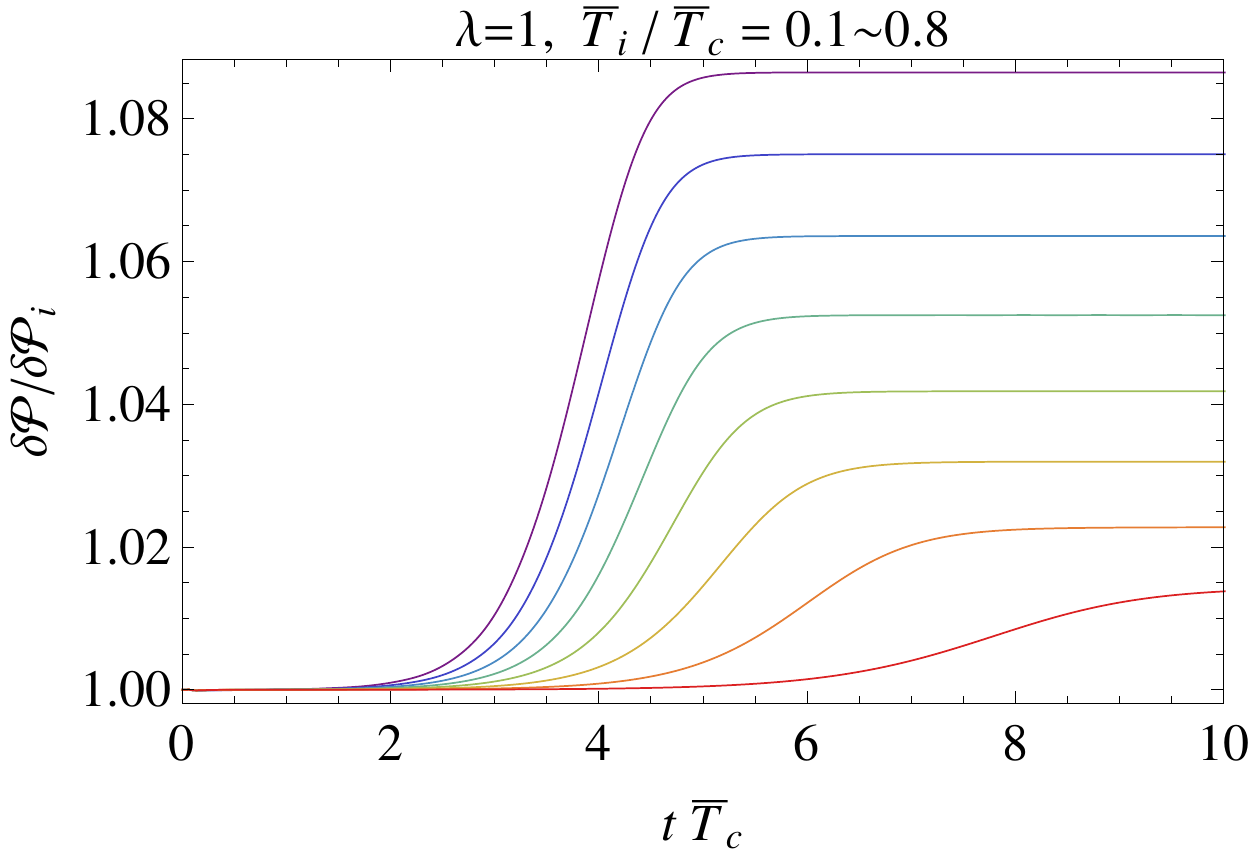}
\hspace{0\textwidth} 
\includegraphics[width =0.48\textwidth]{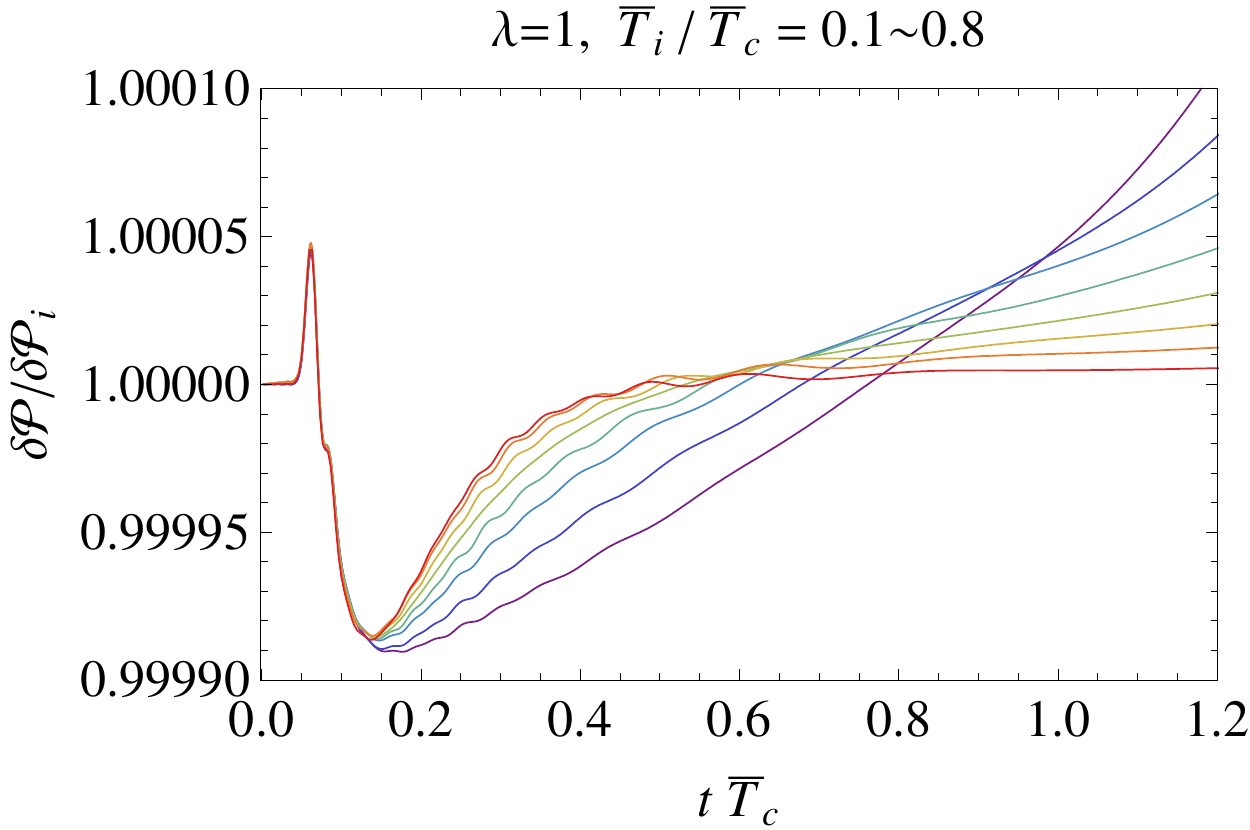} 
\caption{Left: The time evolution of anisotropic pressure $\dP$ for $q=2$ and $\lambda = 1$, normalized by initial pressure $\dP_i$ . The solid curves from top to bottom correspond to $\bT_i/\bT_c = 0.1\sim 0.8$ with increment 0.1. Right: The early time response of the anisotropic pressure to the scalar field perturbation. The according color indicates the same initial temperature. }\label{dPbelowTc}
\end{figure}

From the solution of bulk fields, we extract the information about boundary operators.\footnote{The initial condition for all results in this subsection is a single wave packet with parameters $a=0.005, \delta = 0.05$ and $z_{\text{max}}=0.3$. } In Figure \ref{O2belowTc}, we show the time evolution of condensate $\boo$ for $q=2$ and $\lambda=1$ with initial temperature $\bT_i/\bT_c$ from $0.1$ to $0.8$. Solid curves are results from nonlinear evolution, and dashed curves are given by $\mathcal{A} \exp\left( -t/t_{\text{relax}} \right)$, where $\mathcal{A} $ is a fitting parameter and $t_{\text{relax}}$ is the relaxation time scale and identified as the inverse of imaginary part of dominant scalar quasinormal modes. They have very good quantitative match before saturation kicks in. We will present more results on quasinormal modes in the next section. The evolution of anisotropic pressure $\dP$ (normalized by initial pressure $\dP_i$) is shown in Figure \ref{dPbelowTc}. $\dP$ also have an exponential growing period but with a different relaxation time scale, which is presumably controlled by quasinormal modes of metric fluctuation.

\begin{figure}[H]
\centering
\includegraphics[width =0.46\textwidth]{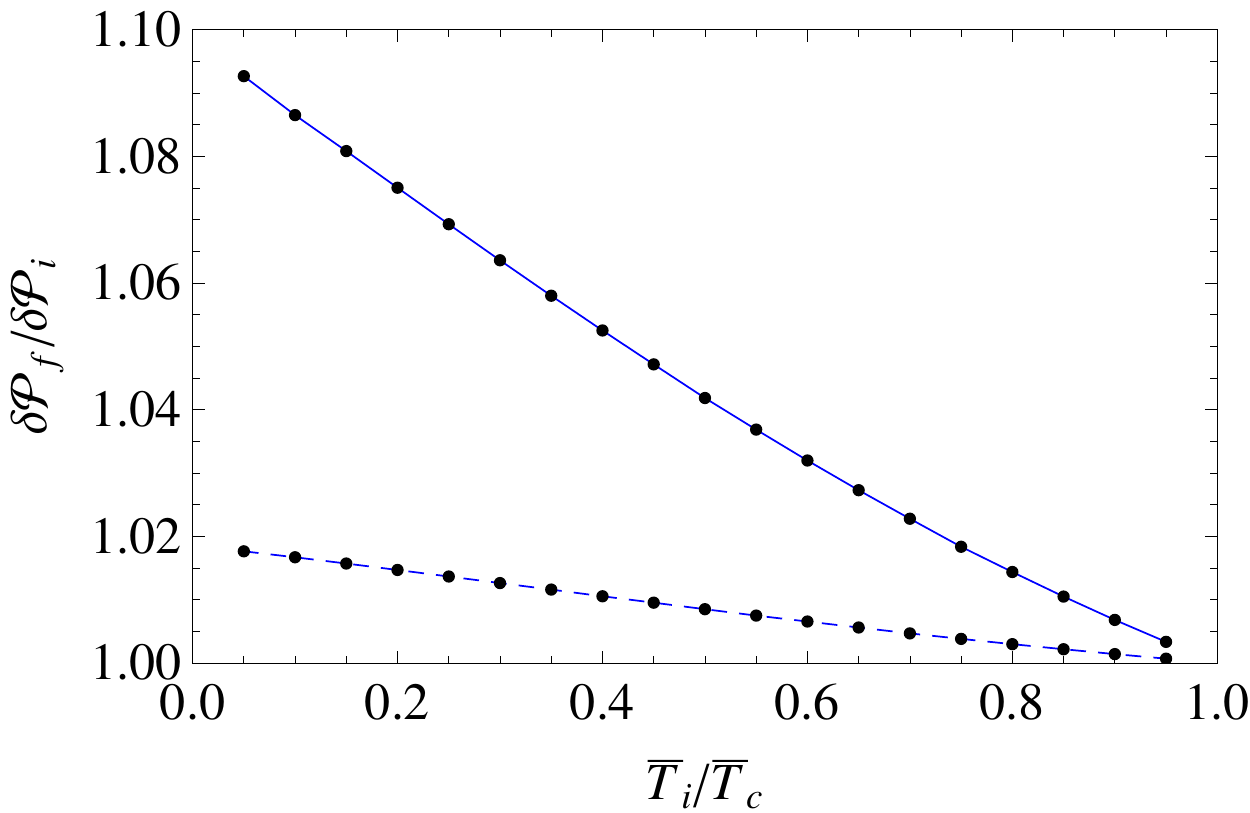}
\hspace{0\textwidth} 
\includegraphics[width =0.46\textwidth]{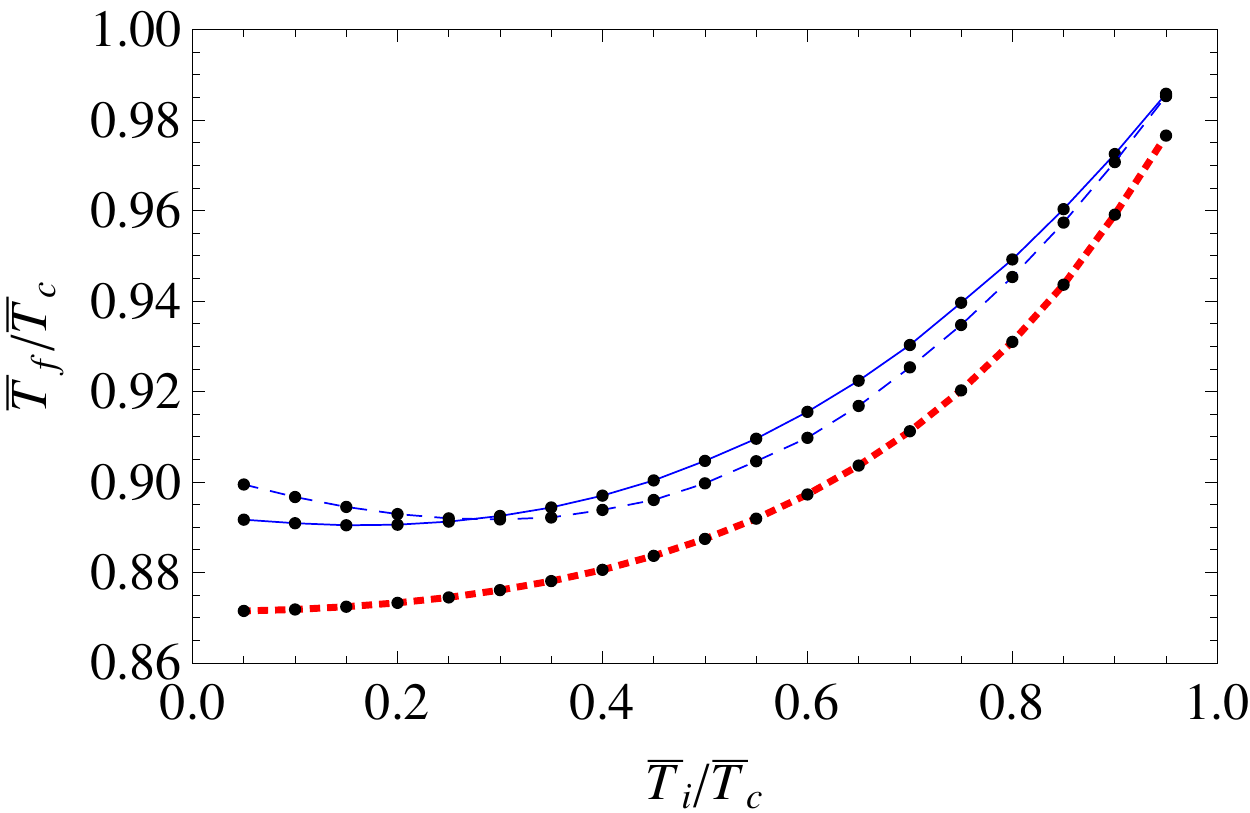} 
\caption{The final anisotropic pressure $\dP_f$ (left) and final temperature $\bT_f$ (right) as a function of initial temperature  $\bT_i$ with $q=2$ for $\lambda =0 $ (red dotted), $\lambda =1 $ (solid) and $\lambda =2$ (dashed). }\label{TivsdPf}
\end{figure}

\begin{figure}[H]
\centering
\includegraphics[width =0.46\textwidth]{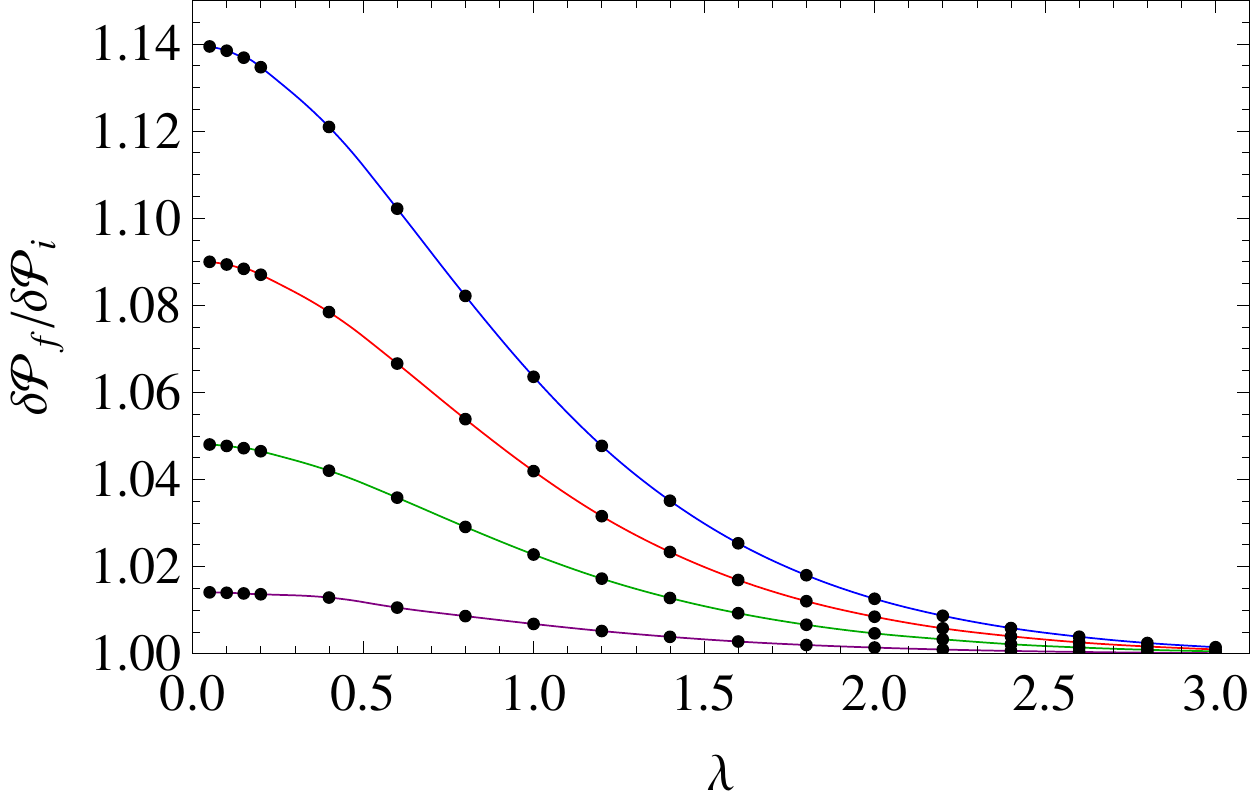} 
\hspace{0\textwidth} 
\includegraphics[width =0.46\textwidth]{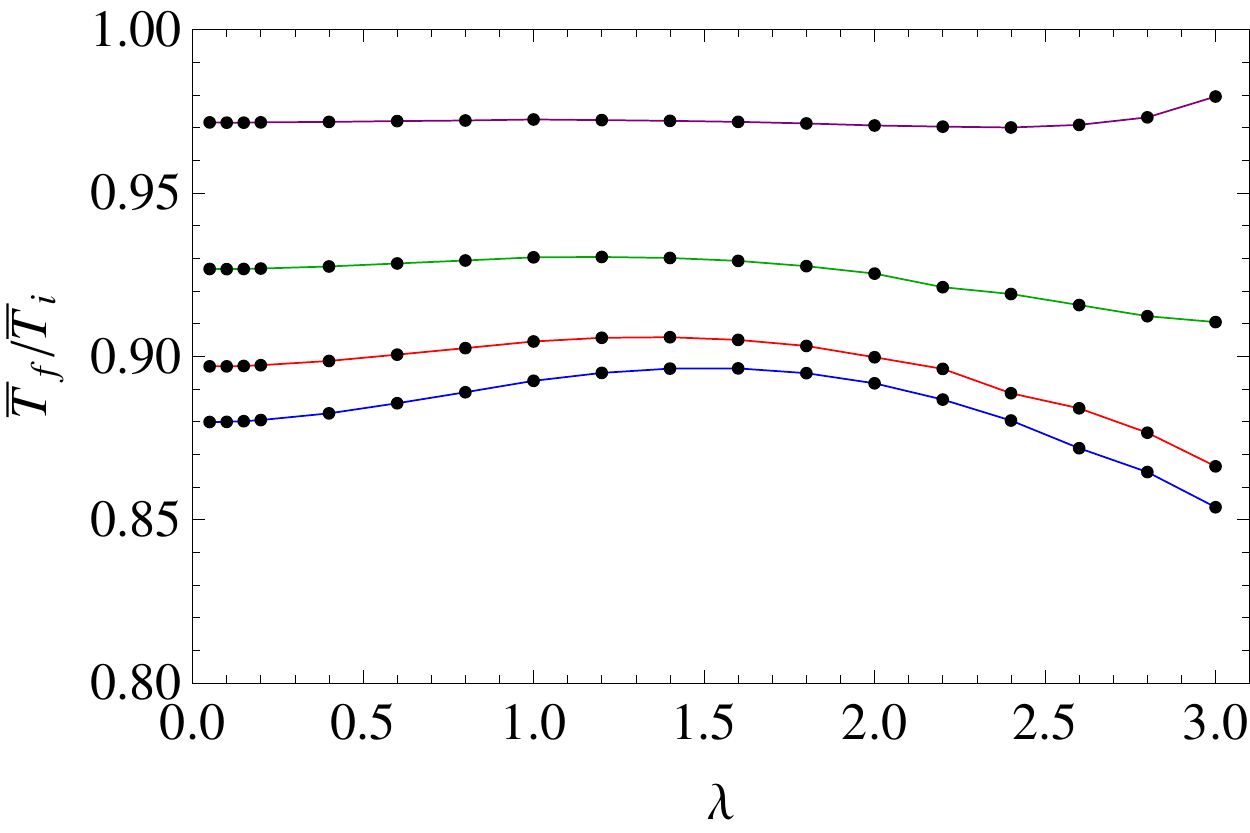} 
\caption{The final anisotropic pressure $\dP_f$ (left) and final temperature $\bT_f$ (right) as a function of anisotropic parameter $\lambda$ with $q=2$ for initial temperature $\bT_i/\bT_c = 0.3,0.5,0.7 $ and $0.9$ from top to bottom (left) and bottom to top (right). The according color indicates the same initial temperature. }\label{lambdavsdPf}
\end{figure}

\begin{figure}[H]
\centering
\includegraphics[width =0.46\textwidth]{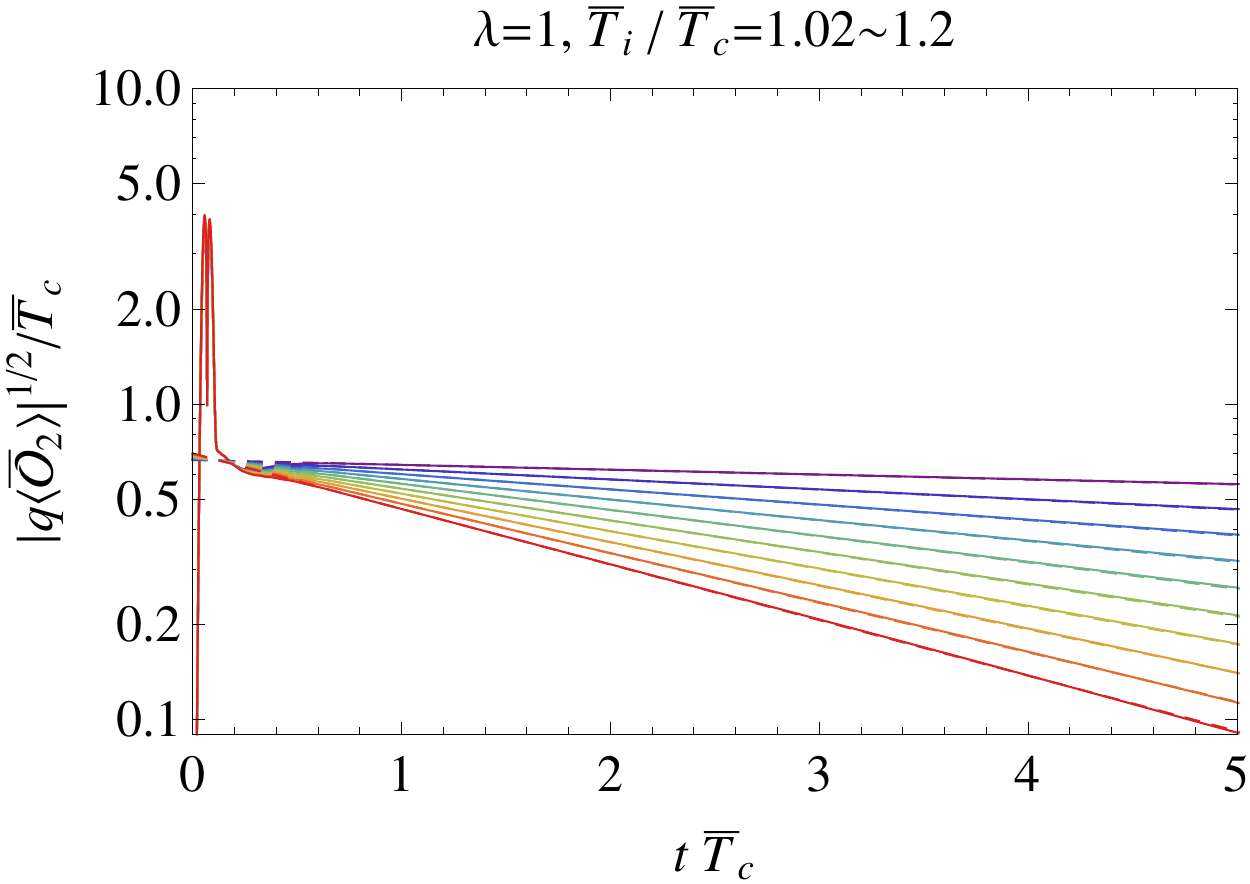} 
\hspace{0\textwidth} 
\includegraphics[width =0.48\textwidth]{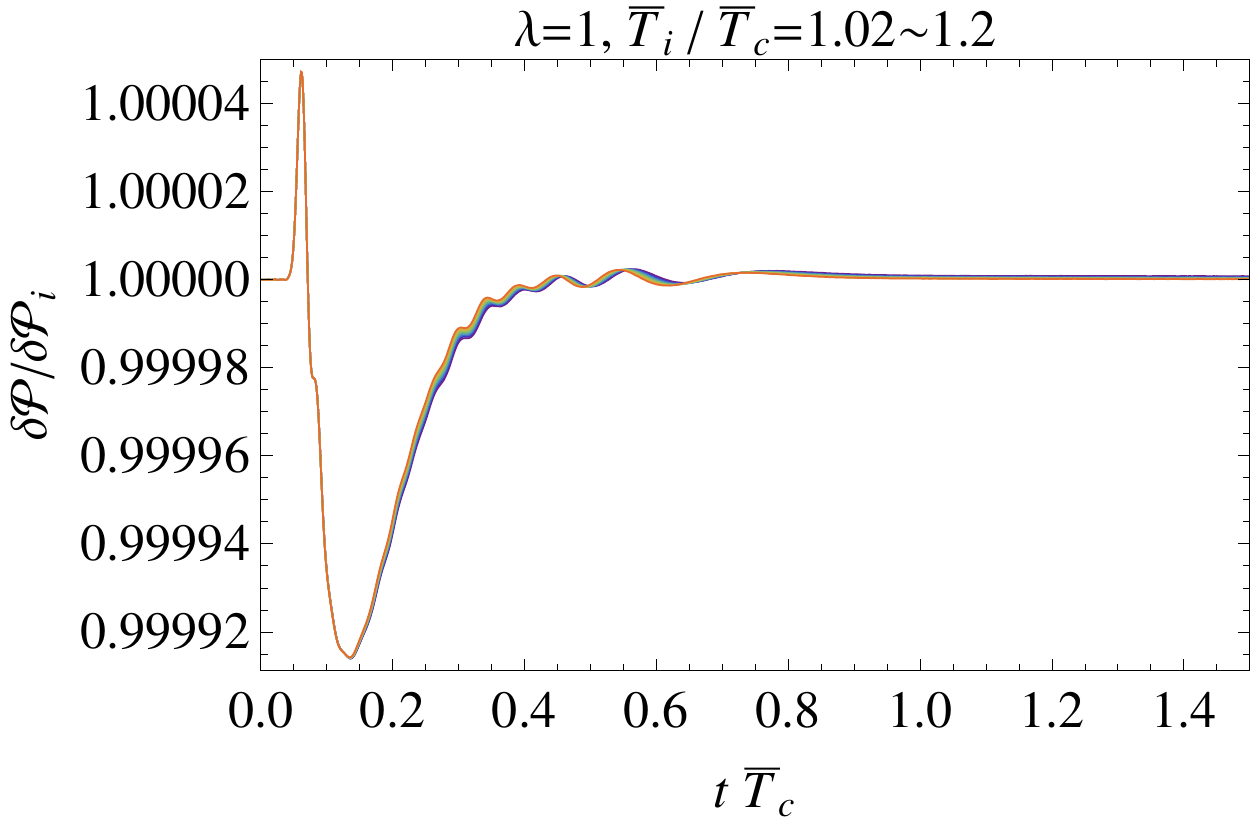} 
\caption{Left: The time evolution of condensate operator for $\lambda = 1$ and $q=2$. The solid curves from top to bottom correspond to $\bT_i/\bT_c = 1.02\sim 1.2$ with increment $0.02$. The dashed curves are relaxation time scale obtained from leading quasinormal modes at corresponding temperature. Right: The response of anisotropic pressure $\dP$. }\label{O2aboveTc}
\end{figure}

It's interesting to see the mapping between initial and final states of the non-equilibrium evolution. In Figure \ref{TivsdPf}, for fixed anisotropic parameter $\lambda$, the pressure ratio $\dP_f/\dP_i$ decreases with initial temperature $\bT_i$ monotonically, while final temperature $\bT_f$ has different behaviour for $\lambda= 1$ and $\lambda =2$. Although the initial temperature may be as low as $0.05~\bT_c$, the relaxation always bring temperature rather close to critical temperature. This feature is also true for the isotropic case (red dotted curve). The superconducting state with low temperature ($\bT_f/\bT_c < 0.5$) seems to be not available in this setup. 
Figure \ref{lambdavsdPf} shows $\dP_f/\dP_i$ and $\bT_f$ as functions of $\lambda$ with initial temperature fixed. Interestingly, the pressure ratios approach constant values as the bulk anisotropy diminishes to zero. 
Figure \ref{O2aboveTc} shows the exponential decaying of initial perturbation for anisotropic black branes with temperature higher than $\bT_c$. The dashed lines are from leading quasinormal modes, which also have a good match with nonlinear evolution. The pressure variation dies out pretty quickly after initial disturbance.


\section{Quasinormal Modes}\label{sec.QNMs}

In the last section, we saw that relaxation time scales extracted from dominant quasinormal modes (QNMs) have very good quantitative agreement with nonlinear evolution in the exponential growing period. We report more QNMs result in this section. QNMs are obtained by solving linearized fluctuation equations subject to infalling boundary condition at horizon and Dirichlet condition at the boundary. For RN-AdS black hole, the scalar field, gauge field and metric perturbation can be separately studied \cite{Policastro:2002se, Miranda:2008vb}, while for hairy black hole one needs to solve coupled fluctuation equations \cite{quasi,wiseman}. For our purposes, it is sufficient to solve decoupled equations and obtain zero momentum sector of scalar QNMs on anisotropic charged black branes. 

The linearized fluctuation equation is 
\begin{align}
&\delta \psi ''(z)+\delta \psi '(z) \left(\frac{F'(z)}{F(z)}+\frac{2 i q \alpha (z)}{F(z)}+\frac{2 i \omega }{F(z)}+\frac{2 \Phi '(z)}{\Phi (z)}\right)\notag\\
&+\delta \psi (z) \left(\frac{i q \left(\Phi (z) \alpha '(z)+2 \alpha (z) \Phi '(z)\right)}{F(z) \Phi (z)}+\frac{2}{z^2 F(z)}+\frac{2 i \omega  \Phi '(z)}{F(z) \Phi (z)}\right)=0\,.\label{linearized}
\end{align}
At horizon ($F(1)=0$), the equation becomes singular. The indicial exponents are $0$ and $-\frac{2i\omega}{F_1}$, corresponding to ingoing and outgoing modes respectively. Althrough $\lambda$ and $B(z)$ do not enter this equation, they affect quasinormal spectrum through coupling with other bulk fields. We employ Chebyshev pseudospectral method and write \eqref{linearized} into matrix form $M \delta\psi = \omega N \delta \psi$. The matrix $M$ and $N$ depend on background fields. This is a generalized eigenvalue problem for $\omega$ and can be solved straightforwardly. Only the modes that are insensitive to the number of Chebyshev grids are reliable.

\begin{figure}[H]
\centering
\includegraphics[width =0.59\textwidth]{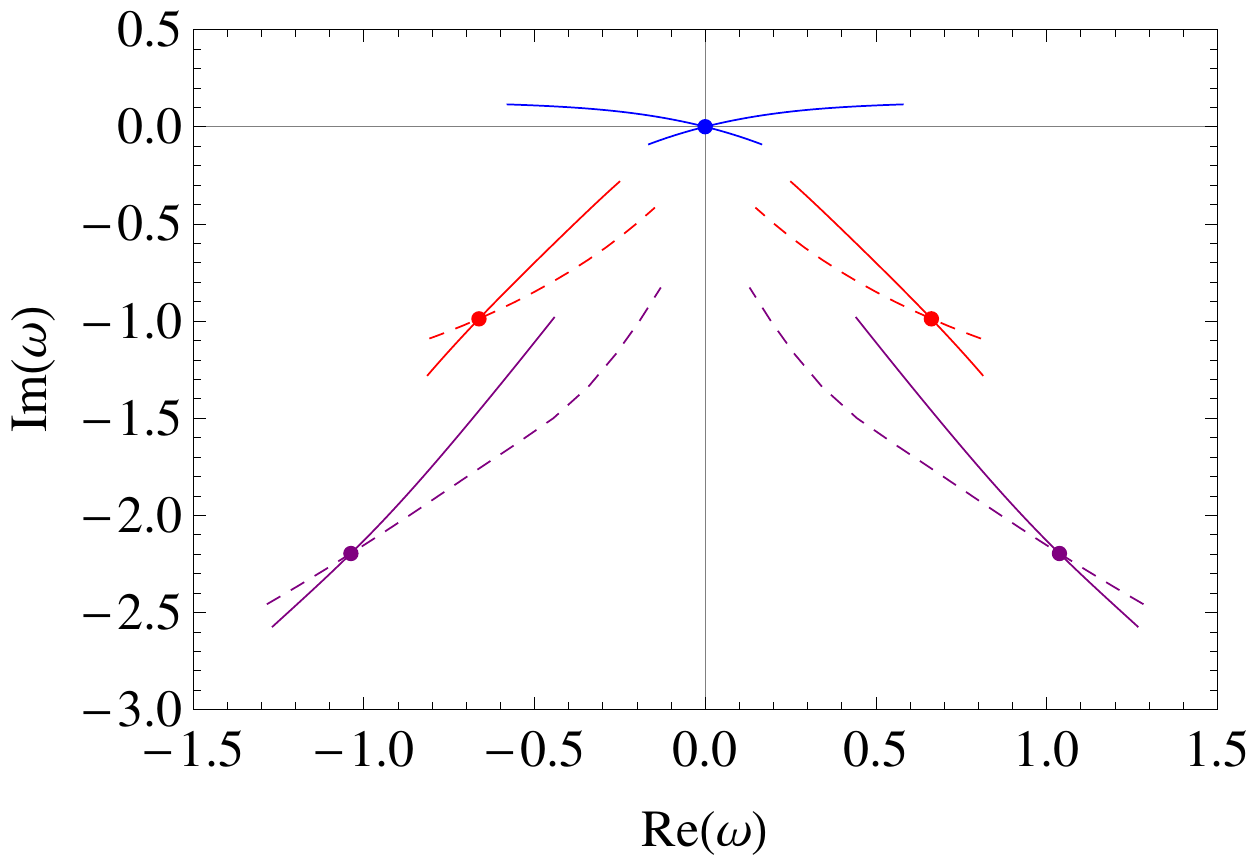} 
\caption{Solid curves are lowest scalar quasinormal modes as function of temperature for $q=2$ and $\lambda=1$ at zero momentum, from $\bT/\bT_c = 1.2$ to $0.5$. The dots indicate the critical temperature points $\bT=\bT_c$. The dashed lines are critical point as function of anisotropic parameter $\lambda$, from $\lambda = 0$ to $3$. }\label{quasisprectum}
\end{figure}

The spectrum for $q=2$ and $\lambda=1$ is shown in Figure \ref{quasisprectum}. The blue, red and purple curves correspond to the first, second and third order of QNMs. As temperature is increased, QNMs ascend in the complex plane. The dots are the critical points, where the dominant mode start to have positive imaginary part indicating the instability of the background. The critical points of subleading QNMs migrate along dashed curves as $\lambda$ is increased from 0 to 3.

\section{Conclusions}

The holographic superconductor model with bulk anisotropy is first introduced in \cite{aniso}. In this work, we study non-equilibrium physics of this model, in particular the dynamical condensation process. In analogy of isotropic case \cite{muruta}, below critical temperature $\bT_c$ we observe a nonlinear evolution from a unstable anisotropic black hole without scalar hair to a stable hairy black hole. This process is identified as a non-equilibrium condensation process in the boundary theory via AdS/CFT correspondence. The condensate operator grows exponentially before saturation. Scalar QNMs are calculated and shown to control the exponential growing behaviour. For $\bT_i>\bT_c$, we observe an exponentially decay in condensate, which matches with QNMs results as well. The holographic renormalization in time-dependent setting is carried out to obtain holographic stress-energy tensor, from which the anisotropic pressure is extracted. It also experiences a similar exponential growth followed by saturation for $\bT_i<\bT_c$ and exponential decay for $\bT_i>\bT_c$, but the relaxation time scale is different from the condensate operator. In addition, we work out the ratio between initial and final anisotropic pressure as a function of anisotropy $\lambda$ for fixed initial temperature. Interestingly, the pressure ratio remains a constant value as $\lambda$ approaches zero.
 
\begin{figure}[H]
\centering
\includegraphics[width =0.47\textwidth]{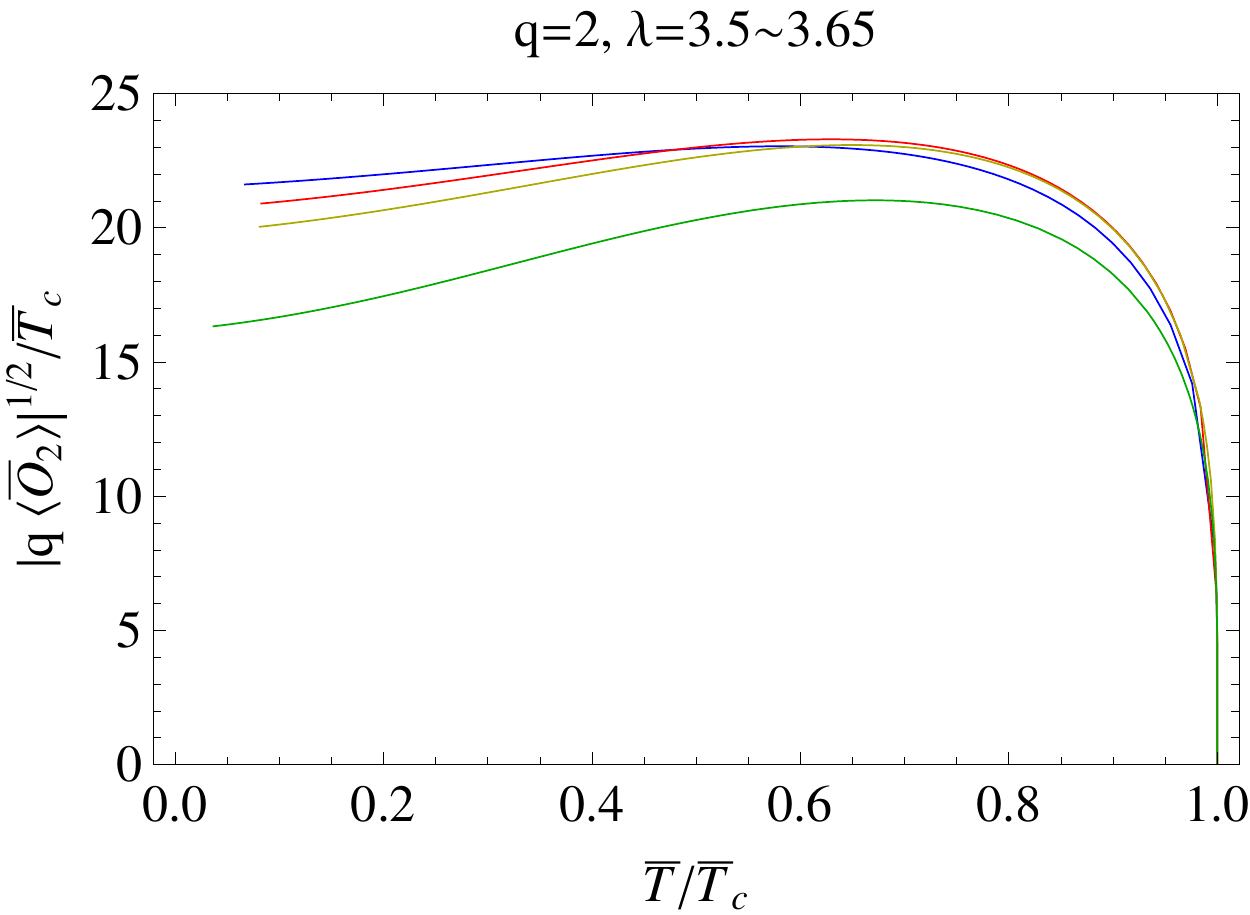} 
\hspace{0\textwidth} 
\includegraphics[width =0.48\textwidth]{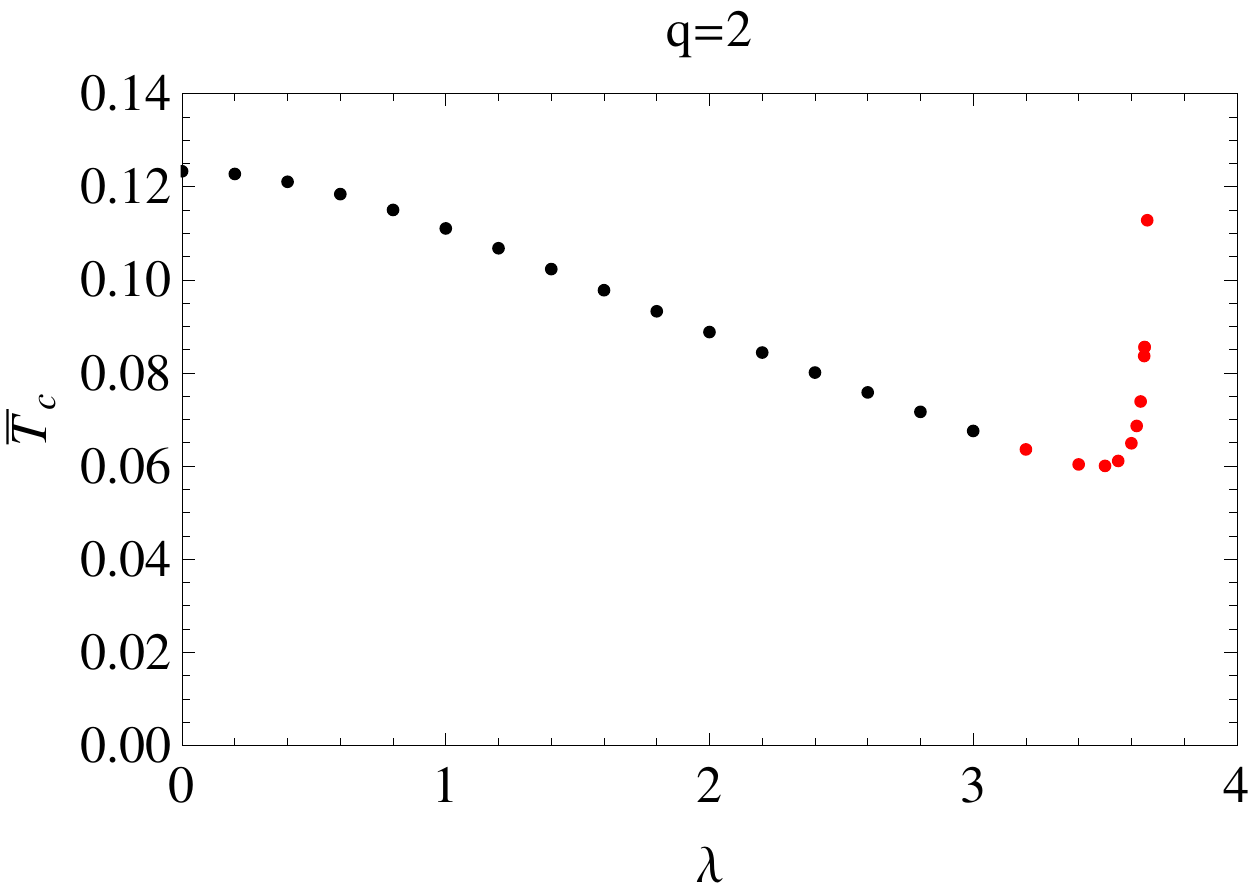} 
\caption{Left: The value of the condensate as a function of temperature with $q = 2$ for $\lambda= 3.5,\,3.55,\,3.60$ and $3.65$, corresponding respectively to the blue, red, yellow and green curves. The condensate gets increasingly stronger suppression at low temperature as $\lambda$ become larger. Right: The critical temperature as a function of anisotropic parameter $\lambda$. Red dots are new data not shown in Figure \protect \ref{equilcondensate}.} \label{large_lambda}
\end{figure}

We close this section with some future directions. For starter, in the equilibrium case of Figure \ref{equilcondensate}, we see that the critical temperature $\bT_c$ decreases with increasing anisotropy during $0<\lambda<3$. Will this trend persist for larger $\lambda$? To our surprise, the preliminary result, Figure \ref{large_lambda}, shows that it's not the case. The critical temperature $\bT_c$ starts to increase around $\lambda = 3.5$ and quickly diverges before reaching $3.7$. The condensate also shows very different behaviour during this range. It gets suppressed at lower temperatures and the suppression becomes stronger as $\lambda$ is increased, in the sharp contrast with the cases of $0<\lambda<3$. The blowing up of the critical temperature is certainly a very interesting phenomenon and implies that, for large anisotropy, hairy solutions exist at arbitrary temperatures and solutions without scalar hair are presumably always unstable. Furthermore, the implication on non-equilibrium physics is not clear at this point. We plan to report more on this subject in the future.
In addition, we only focus on studying time evolution of local one-point functions during condensation process in this work. We will turn our attention to non-local holographic entanglement entropy probe in later investigation. It would also be very interesting to study the quantum quench of the anisotropic system. The last but not least, the dilaton field is taken to be static and decoupled from other matter fields. It's intriguing to introduce a coupling with other fields and achieve the anisotropy in a more dynamical way.

\acknowledgments
We would like to thank Shunichiro Kinoshita, Li Li, Julian Sonner, Norihiro Tanahashi, Chanyong Park, Toby Wiseman and Yang Zhou for helpful correspondence and discussion. XJB thanks Hai-Qing Zhang for teaching him the Chebyshev method and many fruitful discussion. This work was supported by the National Research Foundation of Korea (NRF) grant funded with grant number 2014R1A2A1A01002306.  M. Park was supported by the National Research Foundation of Korea (NRF) funded by the Korea government with the grant No. 2013R1A6A3A01065975.


\bibliography{bibfile}
\bibliographystyle{JHEP}

\end{document}